\documentclass[12pt]{article}

\usepackage{newtxtext,newtxmath}

\usepackage{graphicx}

\usepackage[letterpaper,margin=1in]{geometry}

\renewenvironment{abstract}
	{\quotation}
	{\endquotation}

\date{}

\makeatletter
\renewcommand{\fnum@figure}{\textbf{Figure \thefigure}}
\renewcommand{\fnum@table}{\textbf{Table \thetable}}
\makeatother

\usepackage{scicite}

\usepackage{url}

\usepackage{amsmath}
\usepackage{graphicx}
\usepackage[table]{xcolor}
\usepackage{booktabs}
\usepackage{makecell}
\usepackage{siunitx}
\usepackage{array}

\def\scititle{\sloppy Escape-Induced Temporally Correlated Noise Driven Universality Crossover}

\title{\bfseries \boldmath \scititle}

\author{Mrinal Manna$^{1\dagger}$, Sourav Mukherjee$^{1\dagger}$, Soumen Giri$^{1}$, Pramod Bhakuni$^{1}$,\\ Sajal Barman$^{1}$,  Arnab Kumar Pariari$^{2}$, Anil Gome$^{1}$, Markus Hucker$^{2}$,\\ V. Raghavendra Reddy$^{1}$, Anupam Roy $^{3}$, Sudipta Roy Barman$^{1}$,  \\ Smarajit Karmakar$^{4}$, Chandana Mondal$^{1}$, Rajib Batabyal$^{1\star}$\\ 
\\
\normalsize{$^{1}$UGC-DAE Consortium for Scientific Research,}\\
\normalsize {Khandwa Road, Indore, 452001, India}\\
\normalsize{$^{2}$Condensed Matter Physics, Weizmann Institute of Science,}\\
\normalsize{Rehovot 7610001, Israel}\\
\normalsize{$^{3}$ Department of Physics, Birla Institute of Technology,}\\
\normalsize{Mesra, Ranchi 835215, India}\\
\normalsize{$^{4}$TIFR Center for Interdisciplinary Science,}\\
\normalsize{Tata Institute of Fundamental Research, Hyderabad 500046, India}\\
\\
\normalsize{$^\dagger$ These authors contributed equally to this work.}\\
\normalsize{$^\ast$ Corresponding author. E-mail: rajibbata@csr.res.in.}\\
}

\newpage
\DeclareUnicodeCharacter{03F5}{\ensuremath{\varepsilon}}
\begin{document} 

\maketitle

\begin{abstract} \bfseries \boldmath
Universal behavior in far-from-equilibrium systems is driven by interactions between transport processes and noise structure. The Kardar–Parisi–Zhang (KPZ) framework predicts that extensions incorporating conserved currents or temporally correlated noise give rise to distinct growth morphologies and universality classes, yet direct experimental realization has remained elusive. Here, we report atomically resolved Sn thin-film growth on Sb-doped MnBi$_2$Te$_4$, revealing a sharp dynamical crossover between two fundamentally different regimes. Early-stage growth follows conserved KPZ scaling, forming two-dimensional islands and stanene layers. Beyond a critical deposition time, temporally correlated noise dominates, driving the nucleation of $\alpha$-Sn clusters, their evolution into faceted grains, and coexistence with faceted $\beta$-Sn. Molecular dynamics simulation and Auger electron spectroscopy show adatom escape as the microscopic origin of temporally correlated noise, providing a microscopic mechanism for the universality crossover. These findings establish, for the first time, that temporal noise correlations can fundamentally alter the scaling class of a growing interface, linking atomistic kinetics to emergent universal behavior.

\end{abstract}

\noindent


Understanding how large-scale order emerges in systems far from equilibrium remains a central challenge in statistical physics. Across diverse driven phenomena—from thin-film growth and bacterial colony expansion to flame fronts and fluid invasion—interfaces evolve through stochastic, kinetically roughened dynamics that self-organize into scale-invariant forms~\cite{Barabi1995, he1992, Maunuksela1997, bonachela2011, matsushita1998, wakita1997, Krug1997, Odor2008, Tauber2014}. In this regime, microscopic details become irrelevant, and the evolution is governed by general features such as dimensionality, conservation laws, symmetries, and noise structure—leading to universal scaling exponents and distinct dynamical universality classes ~\cite{Barabi1995, Krug1997, Hohenberg1977, Wilson1974, Odor2008,Tauber2014, Ramasco2000}.

Among these, the Kardar–Parisi–Zhang (KPZ) framework~\cite{KPZ1986} has emerged as the paradigmatic description of nonequilibrium surface growth, capturing the interplay between nonlinear growth, noise, and relaxation. Its various extensions—incorporating conserved surface currents~\cite{LaiSarma1991, SunGuoGrant1989}, spatial or temporal noise correlations ~\cite{Halpin-Healy1989, Medina1989, PMeakin_1989, LamSander1992, Katzav2004, Strack2015, Song_2016, AlejandroLopez2019, Song2021, Song_2021,WangXia2025} —span a rich landscape of universality classes that describe real interfaces. Among them, the conserved KPZ (cKPZ) class models ideal molecular-beam epitaxy (MBE), where surface diffusion dominates and mass is conserved, while temporally correlated noise in the KPZ (TCN-KPZ) equation profoundly modifies interfacial dynamics, producing anomalous roughening, clustering, and faceted morphologies~\cite{AlejandroLopez2019, Song_2021, Song2021}. Despite extensive theoretical advances predicting the effects of temporally correlated noise within the KPZ framework, a direct experimental observation of the dynamical crossover from cKPZ to TCN-KPZ — together with a microscopic identification of the origin of such correlations — has remained elusive, leaving a crucial gap in our understanding of how atomistic kinetics govern emergent universality in nonequilibrium surface growth.

Here, we report the first experimental realization of a universality crossover from the cKPZ class to a TCN–KPZ regime during nonequilibrium Sn thin-film growth on $30\%$ Sb-doped MnBi$_2$Te$_4$ (MBST). By combining atomically resolved scanning tunneling microscopy (STM), scaling analysis, and molecular dynamics (MD) simulations, we uncover a striking transition at growth time, $t = 5.4$~min—from two-dimensional island and stanene-layer growth obeying cKPZ scaling to a late-time regime characterized by clustered mounds and faceted grains governed by temporally correlated noise. The extracted exponents quantitatively match theoretical predictions for cKPZ and TCN–KPZ classes, respectively, confirming a dynamical universality crossover. Simulations reproduce this transition and reveal its microscopic origin: the spontaneous escape of adatoms during growth generates temporally correlated fluctuations in the effective noise term. These escape-induced noise correlations—power-law at short times and exponentially decaying at longer scales—drive the interface from conserved to TCN-KPZ behavior.

Our results establish a direct microscopic mechanism—escape-induced temporally correlated noise—that bridges two nonequilibrium universality classes, connecting atomistic kinetics to emergent scaling laws. This finding not only confirms experimentally a long-standing theoretical prediction~\cite{Song_2021, WangXia2025} but also provides a predictive framework for controlling non-equilibrium morphology through noise correlations in (2+1) dimension.

\begin{figure} 
	\centering
	\includegraphics[width=1\textwidth]{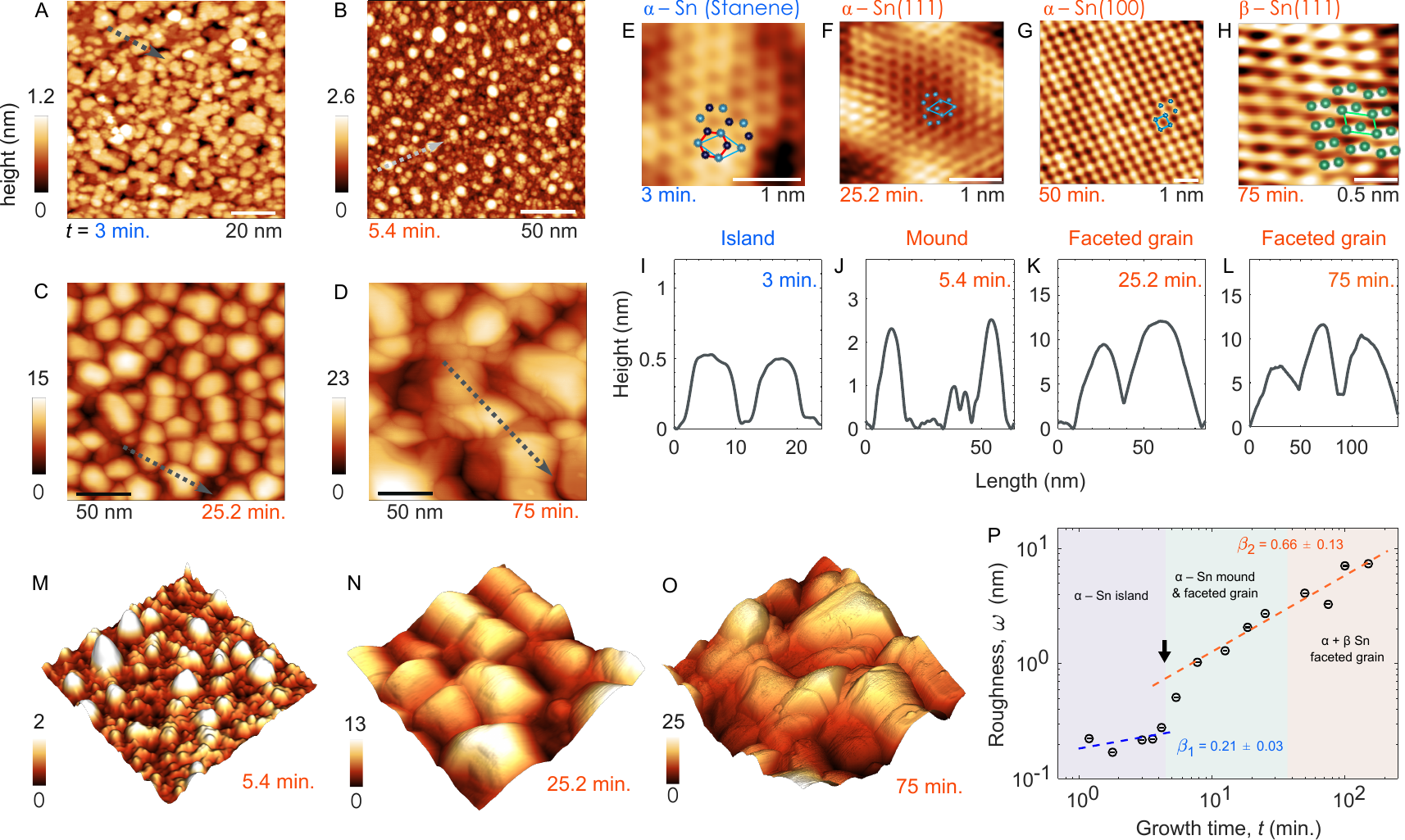} 

	\caption{\textbf{Universality crossover in interface growth.}
		(\textbf{A-D}) STM topographies showing the sequence of growth morphologies with increasing deposition time: (\textbf{A}) early-stage alloyed and few-layer stanene islands, (\textbf{B}) abrupt onset of clustering at $t = 5.4$~min, (\textbf{C}) coalesced mound-like structures, and (\textbf{D}) faceted grains at later stages.(\textbf{E–H}) Atomically resolved STM images revealing (\textbf{E}) buckled stanene, (\textbf{F}) single-phase $\alpha$-Sn grain with (111) facets, and (\textbf{G–H}) coexistence of $\alpha$-Sn grain with (100) and $\beta$-Sn grain with (111) facets at late times. (\textbf{I–L}) height profiles along the marked arrows in (A-D) and 3D view of topographies (\textbf{M–O}) emphasize the morphological transition from 2D islands to 3D faceted grains. (\textbf{P}) Root-mean-square surface roughness, $\omega(t)$, extracted from STM height maps, exhibits two distinct power-law regimes (transition marked by arrow).}
	\label{fig:1} 
\end{figure}

Sn growth on MBST provides a versatile platform to explore a spectrum of growth morphologies across multiple crystalline phases, captured in real time via scanning tunneling microscopy (STM) over extended deposition time (fig.~\ref{figs1}). At early stages of growth, the Sn interface forms two-dimensional (2D) islands atop MBST. In this regime, strong surface alloying occurs: two layers of Sn intermix with the topmost Te and Bi/Sb layers of the substrate, forming a buffer layer~\cite{sajal2024}. Upon continued deposition, few-layer stanene islands nucleate and grow on this alloyed surface (Fig.~\ref{fig:1}A, I). By $t = 4.2$\, min, up to three stanene layers are observed, accompanied by a marked decrease in lateral coverage (fig.~\ref{figs2}). Atomically resolved STM imaging of stanene (Fig.~\ref{fig:1}E) reveals a surface lattice constant of 0.46\, nm—approximately $\sim$5\% larger than that of the substrate ($\sim$0.44\, nm). LEED analysis confirms that this enhanced lattice constant originates in the buffer layer and persists across successive stanene layers ~\cite{sajal2024}. A larger lattice constant in the buffer layer signifies substantial strain relaxation through chemical interactions, providing a favorable template for low-strain stanene growth due to the small residual lattice mismatch of $\sim$~1.5~\% with respect to the free standing stanene. Contrary to this expectation, stanene does not form a continuous monolayer but instead nucleates as islands and evolves into bilayers atop the first layer, leaving the buffer only partially covered. With continued growth, the first-layer stanene eventually achieves full coverage of the buffer, while the lateral extent of the bilayer stanene also increases. The unusual growth behavior stems from the intrinsic bonding of $\alpha$-Sn. Unlike graphene stabilized by planar \( sp^{2} \) bonding, stanene favors an \( sp^{3} \)-like hybridization due to the larger ionic radius of Sn, yielding a buckled lattice rather than a flat honeycomb. The bond angle, \( \theta = 111.4^{\circ} \), corresponds to an effective hybridization index \( sp^{D} \) with \( D = -1/\cos{\theta} \approx 2.74 \), larger than in germanene (2.63), silicene (2.27), or graphene (2.00). This enhanced \( sp^{3} \) character makes stanene intrinsically less stable as a two-dimensional material, with dangling bonds in the buckled lattice promoting a transition toward 3D clustering ~\cite{Xu2018, cahangirov2016}. However, due to residual hydrogen atoms in the ultra-high vacuum chamber, H-passivated stanene films exhibit enhanced stability compared to their bare counterparts ~\cite{Zhao2022, sajal2024}, facilitating the initial formation of stanene layer.

This is exactly what happens at $t = 5.4$\,~min, the morphology undergoes an abrupt transition: Sn no longer grows as extended stanene layers but instead forms randomly distributed clusters (Fig.~\ref{fig:1}B, J). With further deposition, these clusters expand both vertically and laterally, coalescing into mound-like structures and eventually forming faceted grains (Fig.~\ref{fig:1}C, D). Atomic-resolution STM reveals that, up to $t = 25.2$\,min, the faceted grains are composed exclusively of the $\alpha$-Sn phase (Fig.~\ref{fig:1}F); at later times, faceted grains exhibiting both $\alpha$- and $\beta$-Sn phases coexist (Fig.~\ref{fig:1}G, H). This fact is also checked from grazing incidence X-ray diffraction experiments (fig.~\ref{figs3}G). Similar phase coexistence in Sn thin films has been reported on various substrates ~\cite{liu2022, ding2023}, even down to the nanowire length scale~\cite{SKhan2023}. These structural transitions are corroborated by height profiles (Fig.~\ref{fig:1}J–L), taken along lines indicated in the STM topographies (Fig.~\ref{fig:1}B–D), and their 3D view (Fig.~\ref{fig:1}M–O), which vividly capture the morphological evolution from 2D islands to faceted grains.

This sequence of morphological transitions as a function of growth time—from 2D alloyed islands and stanene layers to mound formation and faceted grains—reflects the complex interplay of nucleation, coalescence, and surface diffusion in non-equilibrium growth, giving rise to kinetic roughening and scale-invariant behavior. This motivates a quantitative analysis of dynamical scaling, characterized by the root-mean-square (rms) surface roughness, \( \omega (t) = \sqrt{ \langle [h(\mathbf{r},t) - \langle h \rangle]^2 \rangle} \), where \( h(\mathbf{r},t) \) is the interface (surface to vaccum) height at position \( \mathbf{r} = (x, y) \) and \( t \). During growth, the roughness evolves as \( \omega(t) \sim t^\beta \), where \( \beta \) is the growth exponent. The measured \( \omega(t) \) as a function of deposition time reveals two distinct power-law regimes (Fig.~\ref{fig:1}P), with growth exponents \( \beta_1 = 0.21 \pm 0.03 \) during the early-stage island growth, and \( \beta_2 = 0.66 \pm 0.13 \) during the later stages of mound and faceted grain formation (onset is marked by black arrow). The rms roughness was computed as the standard deviation of the STM topography height map, providing an unbiased estimate independent of scanning direction or instrumental anisotropy. This indicates a dynamical crossover between two fundamentally different growth modes. The physical processes underlying these growth modes can be quantitatively characterized through critical scaling exponents: the roughness exponent \( (\alpha )\), the dynamic exponent \(( z )\) (share the scaling relation \( z = \alpha / \beta \)), and the growth exponent \( \beta \), which together describe the surface across spatiotemporal scales pinning down the universality class and growth modes  ~\cite{Ramasco2000, Karabacak2001, Casiraghi2003, Almeida2014, Durr2003, Drotar2000, Cordoba2009}.

\begin{figure} 
	\centering
	\includegraphics[width=1\textwidth]{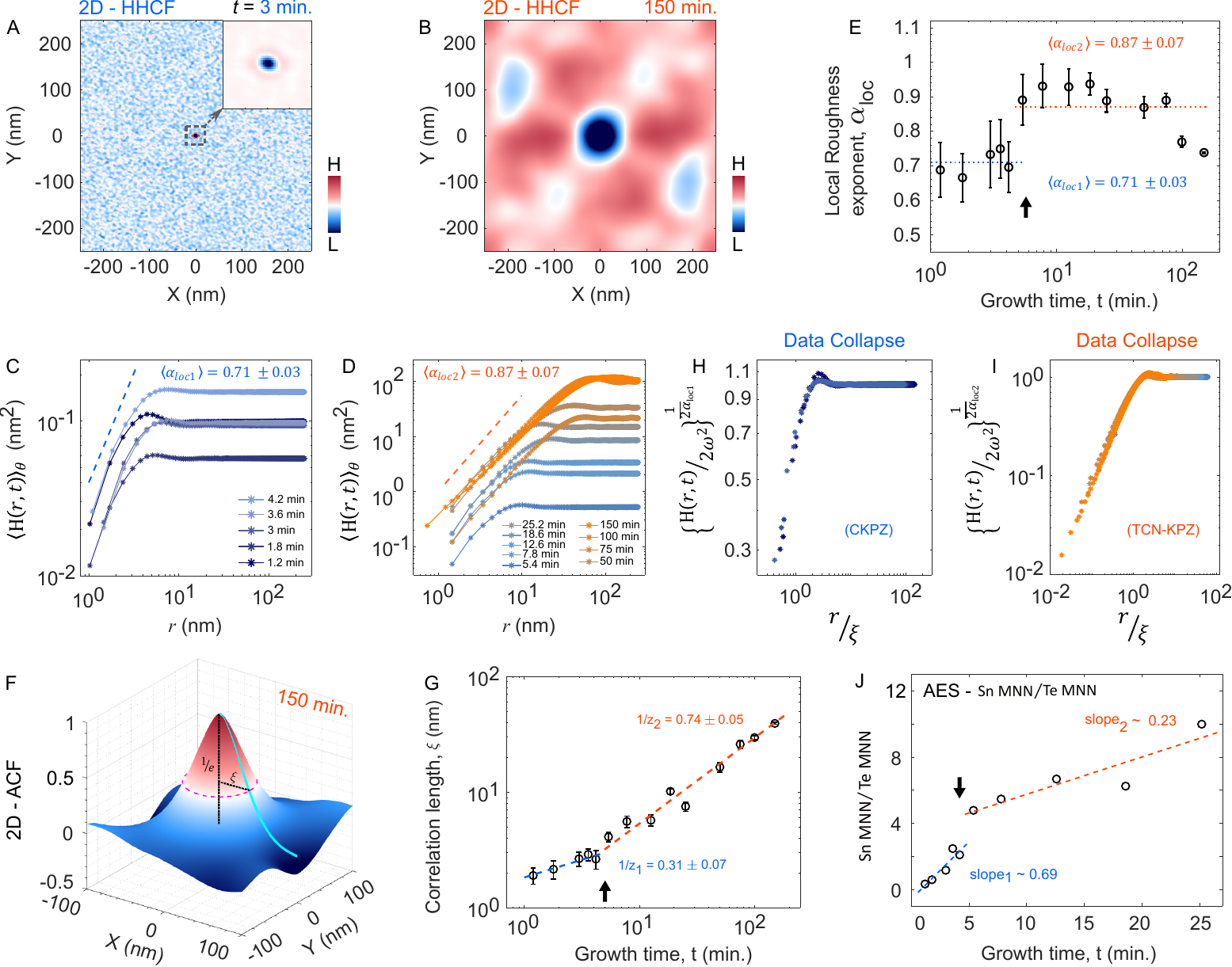} 

	\caption{\textbf{Conserved KPZ and temporally correlated noise KPZ classes and their scaling}
		\textbf{(A, B)} Two-dimensional height–height correlation functions $H(\mathbf{r},t)$ from early (island) and late (mound/faceted) regimes. \textbf{(C, D)} Power-law fits of angularly averaged $\langle H(r,t)\rangle_\theta$ yield local roughness exponents $\alpha_{\mathrm{loc1}}$ and $\alpha_{\mathrm{loc2}}$. \textbf{(E)} Temporal evolution of $\alpha_{\mathrm{loc}}$ exhibits a sharp crossover at $t = 5.4$~min. \textbf{(F)} 2D autocorrelation function and definition of lateral correlation length. \textbf{(G)} Lateral correlation length $\xi(t) \sim t^{1/z}$ yields $1/z_{1} $ and $1/z_{2} $. \textbf{(H, I)} Rescaled $H(\mathbf{r},t)$ collapses onto universal curves for each regime. \textbf{(J)} AES intensity ratio Sn/Te versus growth time shows a reduction in Sn sticking probability from $t = 5.4$~min.}
	\label{fig:2} 
\end{figure}

The height–height correlation function is defined as \( H(\mathbf{r}, t) = \langle [h(\mathbf{r}_0 + \mathbf{r}, t) - h(\mathbf{r}_0, t)]^2 \rangle \), which measures the mean-squared height difference between two surface points separated by a distance \( r = |\mathbf{r}| \). For \( r \ll \xi(t) \), it exhibits power-law scaling as \( H(r, t) \sim \rho^2 r^{2\alpha_{\mathrm{loc}}} \), where \( \rho \) is the average local slope, \( \alpha_{\mathrm{loc}} \) is the local roughness exponent, and \( \xi(t) \) is the lateral correlation length. For \( r \gg \xi(t) \), the correlation saturates as \( H(r, t) \sim 2\omega^2 \)~\cite{Ramasco2000, Cordoba2009, AlejandroLopez2019, Song_2021, Zhao2000}. We computed the two-dimensional \( H(\mathbf{r}, t) \) for all data sets (fig.~\ref{figs4}), and present representative examples from the early-stage island regime and the late-stage mound/faceted-grain regime in Fig.~\ref{fig:2}A and B. To extract \( \alpha_{\mathrm{loc}} \), we performed power-law fits to the angularly averaged height–height correlation function, \( \langle H(r, t) \rangle_\theta \), as shown in Fig.~\ref{fig:2}C and D (see fig.~\ref{figs5} for individual fits). The resulting local roughness exponent \( \alpha_{\mathrm{loc}} \) as a function of growth time is plotted in Fig.~\ref{fig:2}E, revealing two distinct regimes with a crossover at \( t = 5.4 \) min (indicated by a black arrow). The averaged values of the local roughness exponent are \( \langle \alpha_{\mathrm{loc1}} \rangle = 0.71 \pm 0.03 \) and \( \langle \alpha_{\mathrm{loc2}} \rangle = 0.87 \pm 0.07 \) for the early and late growth regimes, respectively. To determine the dynamic exponent \( z \), we analyzed the time evolution of the lateral correlation length \( \xi(t) \sim t^{1/z} \). This was extracted from the two-dimensional autocorrelation function \( C(\mathbf{r}, t) = \langle h(\mathbf{r}_0, t)\, h(\mathbf{r}_0 + \mathbf{r}, t) \rangle \) (fig.~\ref{figs6}), which quantifies how height fluctuations at two points separated by \( r \) are correlated. The correlation length \( \xi(t) \) was defined as the distance at which the angularly averaged correlation function \( \langle C(r, t) \rangle_\theta \) decays to \( 1/e \) of its maximum (fig.~\ref{figs7}), and is plotted in Fig.~\ref{fig:2}F. A clear power-law scaling of \( \xi(t) \) is observed in both regimes with the exponents \( 1/z_{\mathrm{1}}  = 0.31 \pm 0.07 \) and \( 1/z_{\mathrm{2}}  = 0.74 \pm 0.05 \) respectively, with the same crossover time at \( t = 5.4 \) min (Fig.~\ref{fig:2}G). All experimentally extracted critical exponents for the two growth regimes are summarized in Table~\ref{tab:1}.

We next compare the critical exponents \( \alpha_{loc} \), \( \beta \), and \( 1/z \), each independently extracted from experimental data in the two growth regimes, without relying on interdependent scaling relations. In the initial regime, these exponents (\( \alpha_{loc1} =0.71 \pm 0.03 \), \( \beta_{1} = 0.21 \pm 0.03\), and \( 1/z_{1} = 0.31 \pm 0.07 \); where global roughness exponent, \( \alpha_{1} = \beta_{1} z_{1} = 0.68 \pm 0.18\) close to the \(\alpha_{loc1}\) within error limit) closely match those predicted by two established models of MBE growth: the Lai-Das Sarma (LD) model~\cite{LaiSarma1991} and the Das Sarma-Tamborenea (DT) model~\cite{SDSarma1991, PunyinduSarma1997}. The LD model describes conserved surface currents in the KPZ framework, mimicking ideal MBE growth with continuous atomic flux and bonding interactions. The DT model assumes random deposition with perfect surface relaxation, where atoms instantly settle into nearest kink sites. Although both models belong to the same asymptotic universality class, they differ in non-asymptotic corrections, particularly in the local slope exponent \( \lambda \), defined by \( \rho(t) \sim t^{\lambda} \), where \( \lambda \neq 0 \) for DT and \( \lambda = 0 \) for LD~\cite{PunyinduSarma2001, Zhao2000}. We also performed power-law fitting of the local slope \( \rho(t) \) to extract the slope exponent \( \lambda \) in both regimes (Figure~\ref{figs8}). In the early growth regime, the data exhibit minimal variation with time and are too scattered to reliably support a power-law, yielding \( \lambda_1 = -0.04 \pm 0.03 \), effectively consistent with \( \lambda_1 = 0 \). This result aligns with the LD model, which predicts a vanishing slope exponent. Furthermore, the DT model predicts mound-like morphologies, which are clearly inconsistent with our observation of island growth. These results support the conclusion that the early-time growth dynamics are governed by the cKPZ framework, specifically within the LD continuum universality class.

In the subsequent growth regime, beyond the crossover time, the surface dynamics exhibit a distinctly different scaling behavior characterized by the extracted critical exponents \( \alpha_{\mathrm{loc2}} = 0.87 \pm 0.07 \), \( \beta_{2} = 0.66 \pm 0.13 \), and \( 1/z_{2} = 0.74 \pm 0.05 \); yielding a global roughness exponent \( \alpha_{2} = \beta_{2} z_{2} = 0.89 \pm 0.19 \), consistent with \(\alpha_{\mathrm{loc2}}\) within uncertainty. These values represent a clear departure from the conserved KPZ scaling of the early regime, indicating the emergence of a different universality class governing the late-time dynamics. Instead, these values closely align with theoretical predictions of the KPZ framework extended to include temporally correlated noise (TCN–KPZ)~\cite{Song_2021}. The introduction of long-range, power-law temporal correlations in the KPZ noise term, $\langle \eta(r,t)\,\eta(r',t') \rangle = 2D \, \delta_{r,r'} \, |t - t'|^{2\theta - 1}$, where $\eta(r,t)$ is the noise, $D$ its amplitude, and $\theta \in [0, \tfrac{1}{2}]$ the correlation exponent, profoundly alters the growth dynamics, yielding modified scaling exponents and macroscopic surface features. Strong correlation (\( \theta \to 0.5 \)) drives the surface to form clusters or mounds, which at longer times evolve into macroscopic faceted patterns that dominate the morphology ~\cite{AlejandroLopez2019,Song_2021,Song2021}. Remarkably, our experimentally extracted exponents are consistent with \(\theta \approx 0.45\) in (2+1)-dimensional TCN-KPZ framework, indicating a crossover from the cKPZ class to a new universality class governed by long-range temporally correlated noise. At $t = 5.4$\,min, the onset of temporally correlated noise manifests as a sharp transition from 2D islands to clustered mounds, which subsequently evolve into faceted grains, as shown in 3D topographies (Figure~\ref{fig:1}M–O) and corresponding height profiles (Figure~\ref{fig:1}I–L), consistent with theoretical predictions for surfaces subject to strongly correlated noise ~\cite{AlejandroLopez2019,Song_2021,Song2021}. To provide a stringent test of universality in the growth dynamics, we examine whether the measured observables collapse onto a single universal curve under appropriate rescaling, as \( r \to \tfrac{r}{\xi} \) and  \( H(\mathbf{r},t) \to \left\{ \tfrac{H(\mathbf{r},t)}{2\omega^{2}} \right\}^{\tfrac{1}{2\alpha_{loc}}} \), applied separately for each region with its respective critical exponents. The resulting data collapse of \(H(\mathbf{r},t)\) across different growth times, shown in Figure~\ref{fig:2}H and I, demonstrates robust convergence, thereby confirming the universality crossover from the cKPZ class to the TCN-KPZ class. The sticking probability of adatoms could be a key parameter, determining the fraction of incident atoms that incorporate into the film versus those that desorb and act as noise in the dynamics. Auger electron spectroscopy (AES) as a function of growth time (Figure~\ref{fig:2}J and Figure~\ref{figs9}) shows a striking change in the Sn/Te intensity ratio: the slope decreases from $\sim 0.7$ in the island-dominated cKPZ regime to $\sim 0.23$ in the mound-to-faceted grain region governed by TCN-KPZ dynamics. This transition, occurring at $t \approx 5.4$ min, coincides precisely with the universality crossover of the critical exponents and the morphological transition, demonstrating a strong suppression of Sn sticking on the surface when temporal noise correlations dominate. This remarkable coincidence of onset time is not a random event; rather, as we discuss later, it marks the precise point at which temporal noise correlations set in.

\begin{figure} 
	\centering
	\includegraphics[width=1\textwidth]{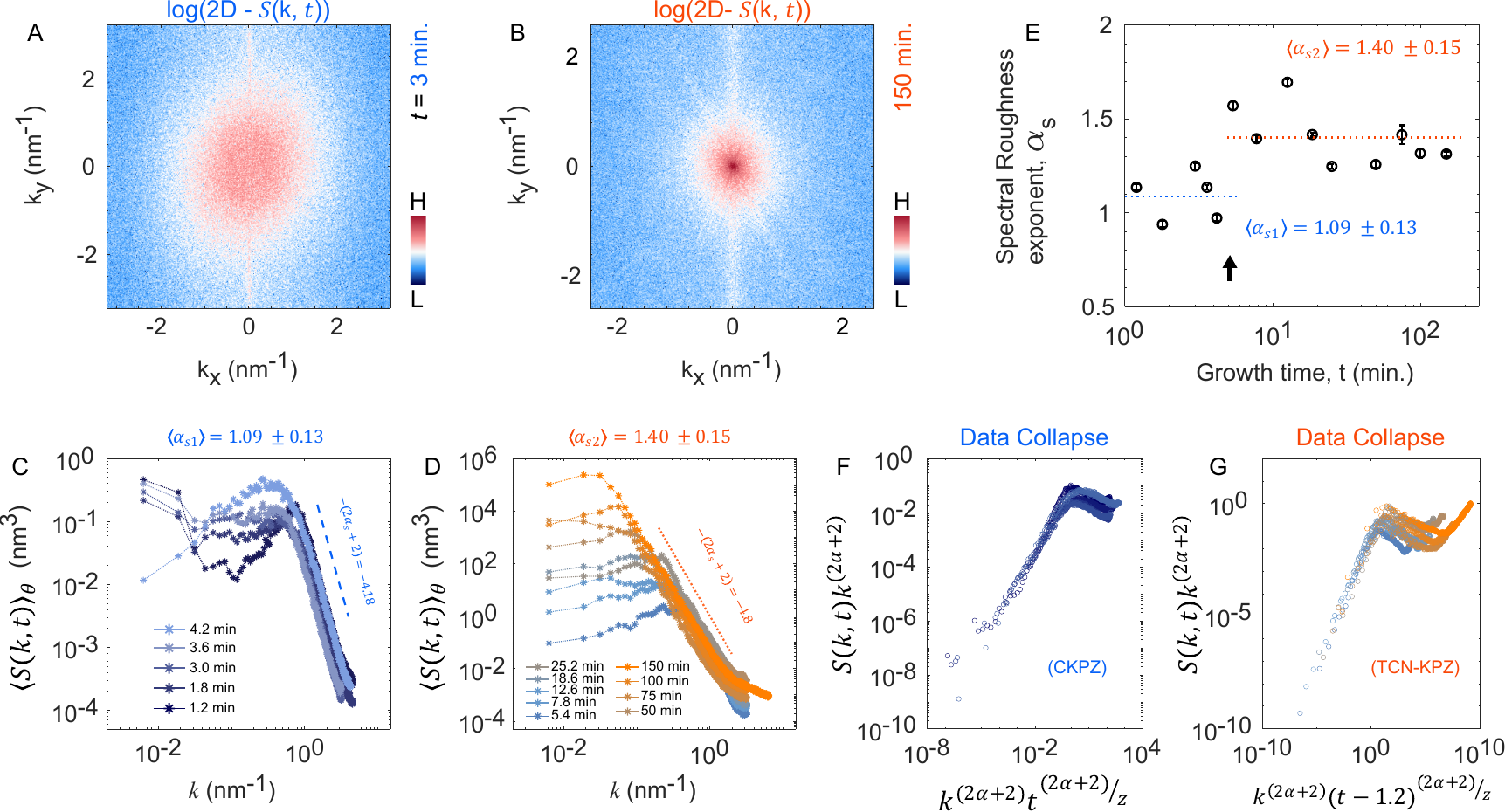} 

	\caption{\textbf{Spectral scaling and emergence of faceted growth driven by temporally correlated noise.} \textbf{(A, B)} Two-dimensional structure factors $S(\mathbf{k},t)$ for island-dominated ($t = 3.0$~min) and mound/faceted ($t = 150$~min) regimes. \textbf{(C, D)} Angularly averaged $\langle S(k,t)\rangle_{\theta}$ exhibit power-law scaling with spectral roughness exponents $\alpha_{s1}$ and $\alpha_{s2}$. \textbf{(E)} Temporal evolution of $\alpha_{s}$. \textbf{(F, G)} Data collapse of rescaled $S(k,t)$.}
	\label{fig:3} 
\end{figure}

Generic kinetic roughening theory predicts that faceted surfaces, owing to their intrinsic geometric morphology, display distinct scaling functions and exponents. A complete characterization of these surfaces is most effectively achieved in Fourier space via the structure factor, \(S(\mathbf{k},t) = \langle {h}(\mathbf{k},t) {h}(-\mathbf{k},t) \rangle,\) where \( \quad {h}(\mathbf{k},t) \equiv (\frac{1}{L})^{d/2} \int dr\, h(\mathbf{r},t) e^{-i \mathbf{k} \cdot \mathbf{r}},\) and $\mathbf{k}$ the wave vector ~\cite{Ramasco2000, AlejandroLopez2019}. Notably, $S(k,t)$ is the key quantity that directly reveals the spectral roughness exponent $\alpha_s$, the defining feature of faceted growing surfaces, where $\alpha_s > 1$ and global roughness $\alpha \neq \alpha_s$ (here, $\alpha_2 \neq \alpha_{s2}$), requiring two independent exponents to fully describe the scaling properties ~\cite{AlejandroLopez2019}. We estimated the 2D $S(k,t)$ for the entire dataset (Figure~\ref{figs10}),with Figure~\ref{fig:3}A and B depicting the island-dominated regime at $t = 3.0$~min and the mound-to-faceted grain regime at $t = 150$~min as illustrative examples. The spectral roughness exponent $\alpha_{s}$ was obtained from power-law fits to the linear scaling regime of the angularly averaged structure factor, $\langle S(k,t) \rangle_{\theta}$ (Figure ~\ref{fig:3}C,D; see Figure ~\ref{figs11} for individual fits). Its temporal evolution (Fig.~\ref{fig:3}E) reveals a sharp crossover at $t = 5.4$~min, separating two scaling regimes. In the early-time regime, we obtain $\langle \alpha_{s1} \rangle = 1.09 \pm 0.13$, for island-dominated growth, whereas at late times  $\langle \alpha_{s2} \rangle = 1.40 \pm 0.15$, satisfies the the condition $\alpha_{s2} > 1$ and global $\alpha_{2} (0.89 \pm 0.19) \neq \alpha_{s2}$ for the onset of faceted morphology. We further performed data collapse of the structure factors by rescaling $k \to k^{(2\alpha+2)} (t - t')^{(2\alpha+2)/z}$ and $S(k,t)\to S(k,t)\,k^{2\alpha+2}$ ~\cite{Luca2007} (Figure~\ref{fig:3}F,G). The most consistent collapse was achieved with the exponents $1/z_{1} = 0.36$, $\alpha_{1} = 0.58$ in the early regime with $t' = 0$, and $1/z_{2} = 0.75$, $\alpha_{2} = 0.88$ in the late regime with effective growth time $t' = 1.2$~min ~\cite{Luca2007}, which are very close to the measured coefficients. These results further provide experimental evidence that temporally correlated noise drives KPZ interfaces into a faceted regime, a phenomenon anticipated by theory and observed in simulations ~\cite{AlejandroLopez2019, Song_2021, Song2021}.

\begin{figure} 
	\centering
	\includegraphics[width=1\textwidth]{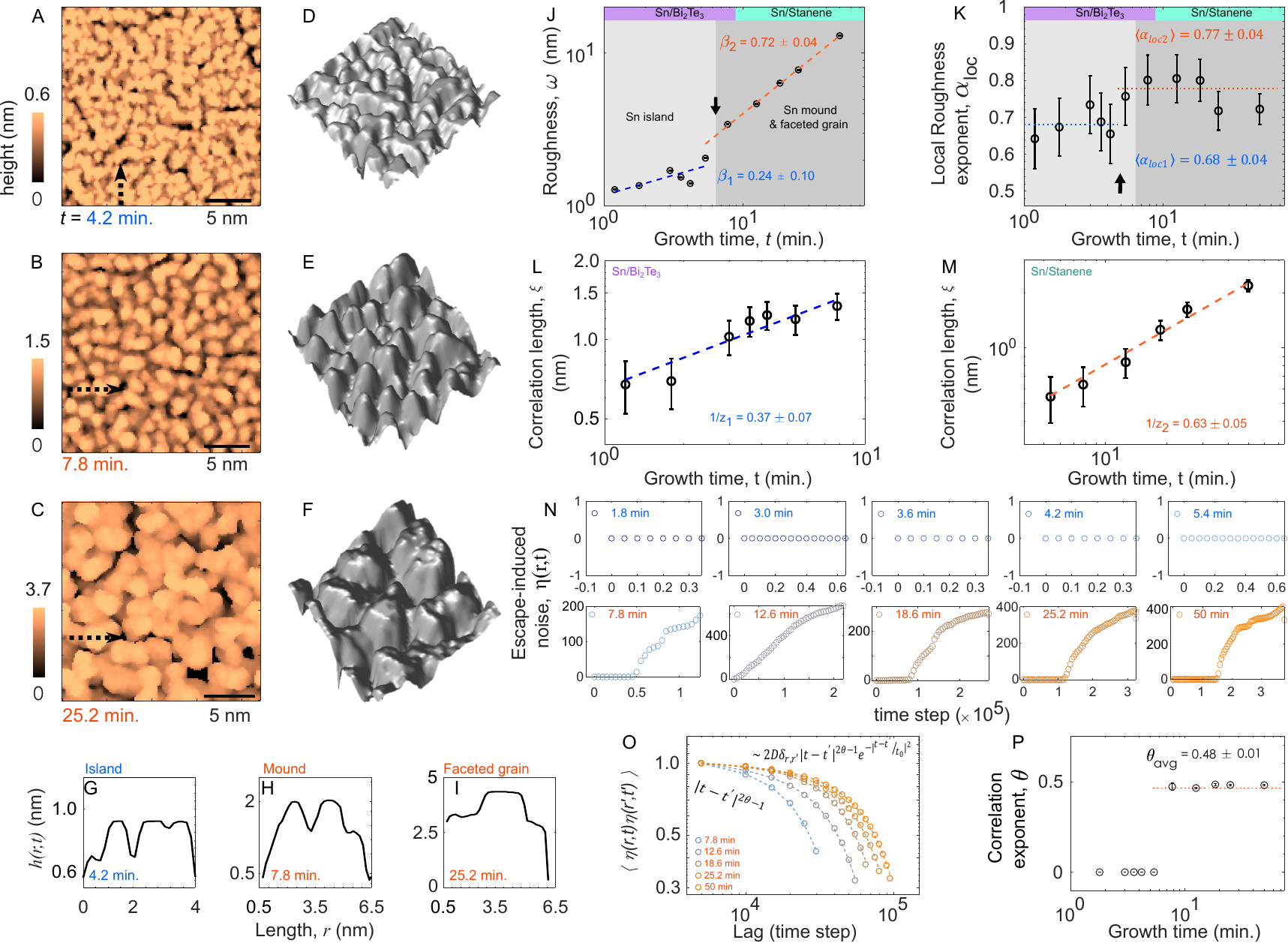} 
	\caption{\textbf{Escape-induced noise and microscopic origin of temporally correlated noise.} \textbf{(A–C)} Simulated Sn surface morphologies showing the transition from island nucleation to mounded and faceted growth. \textbf{(D–F)} 3D renderings corresponding to \textbf{(A–C)} and \textbf{(G–I)} height profiles along the arrows marked in (A-C). \textbf{(J)} Temporal evolution of the rms roughness $\omega(t)$ exhibits two distinct power-law regimes. \textbf{(K–M)} Local roughness and dynamic exponents $\alpha_{\mathrm{loc}}$ and $z$ show a crossover at $t = 5.4$~min. \textbf{(N)} Particle-tracking analysis during deposition showing that up to $t = 5.4$~min all deposited atoms are incorporated into growth (top), whereas at later times a significant fraction escape the surface (bottom), generating temporally correlated noise. \textbf{(O)} Temporal correlations of the escape-induced noise exhibit a hybrid form - power-law at short lags and exponential decay at longer times. \textbf{(P)} The average over the fitted correlation exponents $\theta_\mathrm{avg} = 0.48 \pm 0.01$ agrees with theory (\,$\theta \approx 0.45$\,), confirming that escape events generate temporally correlated noise.}
	\label{fig:4} 
\end{figure}

To probe the universality crossover from cKPZ to TCN-KPZ and uncover the origin of temporally correlated noise, we carried out extensive molecular dynamics simulations of non-equilibrium Sn growth. Interestingly, reproducing the experimental crossover required two substrates: a Bi$_2$Te$_3$ slab as an analog of MBST, which lacks reliable potentials for its Mn and Sb dopants yet exhibits the same buffer-layer formation before stanene~\cite{sajal2024, li2020}, and a buckled honeycomb Sn layer representing stanene for the late-time regime (growth movies are in SM ~\cite{methods}). This combination exactly reproduce the experimental observation here. Figure~\ref{fig:4}A–C display the simulated surface evolution (Figure~\ref{figs12} for entire regime), showing an initial island nucleation that progressively transitions to clustered and mounded morphologies, eventually developing into faceted grains. To highlight the morphological evolution, corresponding area-selected 3D topographies (Figure~\ref{fig:4}D–F) and representative height profiles along the marked arrows in Figure~\ref{fig:4}A–C are presented (Figure~\ref{fig:4}G–I), closely reproducing the experimental observations (Figure~\ref{fig:1}). On these simulated topographies, we performed similar analyses to determine the critical exponents across the entire growth regime. The simulated rms roughness, \( \omega(t) \), as a function of deposition time exhibits two distinct power-law regimes (Figure~\ref{fig:4}J), mirroring the experimental data (Figure~\ref{fig:1}P). The growth exponents were found to be \( \beta_1 = 0.24 \pm 0.10 \) during the early-stage island nucleation and \( \beta_2 = 0.72 \pm 0.04 \) during the subsequent mound and faceted grain growth, with the onset of the crossover indicated by a black arrow. Likewise, the temporal evolution of the local roughness exponent \( \alpha_{\mathrm{loc}} \) (Figure~\ref{fig:4}K) and dynamic exponent \( z \) (Figure~\ref{fig:4}L–M) reveals two clearly separated regimes, characterized by \( \alpha_{\mathrm{loc1}} = 0.68 \pm 0.04 \) and \( 1/z_{1} = 0.37 \pm 0.07 \) for the early-stage growth, and \( \alpha_{\mathrm{loc2}} = 0.77 \pm 0.04 \) and \( 1/z_{2} = 0.63 \pm 0.05 \) for the late-stage morphology evolution. All simulated exponents exhibit excellent agreement (within uncertainties, Table~\ref{tab:1}) with the experimentally determined values for the cKPZ and TCN-KPZ regimes, and the corresponding data collapse demonstrates remarkable convergence (Figure~\ref{figs13}). During the simulations, we observed a notable interplay between Sn growth and the underlying substrates. On the Bi$_2$Te$_3$ slab, the universality crossover is clearly captured near $t \sim 5.4$~min, in agreement with experimental observations, however, continued growth on this substrate beyond $t = 7.8$~min leads to a deviation of the critical exponents that define the universality class. Examining the experimentally observed stanene island morphologies near the transition time $t = 5.4$~min reveals that the first-layer stanene fully covers the underlying buffer (Figure~\ref{figs2}B–D), before clusters begin to form, marking the onset of the TCN-KPZ regime. Notably, in simulations on Bi$_2$Te$_3$, a fully covered stanene layer is never achieved. Guided by experimental observations, we switched the substrate to stanene, and subsequent growth reproduces the experimentally measured exponents with remarkable consistency. A possible reason that simulations on Bi$_2$Te$_3$ unable to capture the late-time behavior is the stabilizing effect of residual hydrogen atoms present in the ultra-high vacuum chamber, which are not included in the simulation.

To uncover the origin of temporally correlated noise, we analyzed the particle dynamics during growth. Remarkably, up to $t = 5.4$~min, all deposited particles are incorporated into islands, contributing fully to growth (top panels of Figure~\ref{fig:4}N). Beyond $t = 7.8$~min, a significant fraction of particles escape the system, giving rise to a distinct escape-induced noise component (bottom panels of Figure~\ref{fig:4}N). The temporal correlations of this noise reveal a complex structure (Figure~\ref{fig:4}O). Rather than a simple power-law decay, the correlations exhibit two regimes: short-time lags follow a power-law, while longer lags decay exponentially. This behavior is accurately described by$\langle \eta(r,t)\,\eta(r',t') \rangle = 2D\,\delta_{r,r'}\,|t - t'|^{2\theta - 1} \exp[-|t - t'|^2/t_0^2]$, where $t_0$ is the characteristic decay time. Fitting this model to the simulation data yields a correlation exponent $\theta_\mathrm{avg} = 0.48 \pm 0.01$ (Figure~\ref{fig:4}P), in close agreement with the theoretical prediction (\(\theta \approx 0.45\)) ~\cite{Song_2021}. Remarkably, this microscopic mechanism is fully consistent with the experimental Auger electron spectroscopy (AES) measurements: the slope of Sn/Te intensity ratio as a function of growth time sharply decreases from $\sim 0.7$ in the island-dominated cKPZ regime to $\sim 0.23$ in the mound-to-faceted grain regime, precisely coinciding with the onset of temporally correlated dynamics. Together, the AES data and simulations demonstrate that the reduction in effective sticking probability arises from escape-induced temporally correlated noise, linking atomistic desorption events to the emergent universality crossover from conserved KPZ to TCN-KPZ behavior.

These findings demonstrate that particle escape during growth acts as a microscopic source of temporally correlated noise, with a dual character: power-law correlations at short times and exponential decay at longer times. This mechanism provides a concrete origin for the crossover from conserved KPZ dynamics to TCN-KPZ behavior, underscoring the fundamental role of adatom escape in shaping interface fluctuations. Early-stage 2D islands and stanene layers follow conserved KPZ scaling, but beyond $t \approx 5.4$~min, long-range temporally correlated noise in the deposition dynamics drive a transition to the temporally correlated TCN-KPZ regime, producing clustered mounds and faceted grains. MD simulations quantitatively reproduce both regimes, linking microscopic particle escape to macroscopic surface universality, and establishing a predictive framework to control non-equilibrium growth and engineer surface morphology in (2+1) dimension.


\renewcommand{\arraystretch}{1.5}
\setlength{\tabcolsep}{2pt}

\begin{table}[ht!]
\centering
\rowcolors{3}{gray!10}{white}
\begin{tabular}{
    |>{\centering\arraybackslash}m{2.2cm}
    |>{\centering\arraybackslash}m{2.2cm}
    |>{\centering\arraybackslash}m{2.2cm}
    |>{\centering\arraybackslash}m{2.2cm}
    |>{\centering\arraybackslash}m{2.5cm}
    |>{\centering\arraybackslash}m{2.2cm}|
}
\hline
\rowcolor{gray!30}
\textbf{Universality Class} & \textbf{(2+1)D} & $\boldsymbol{\alpha_{\text{loc}}}$ & $\boldsymbol{\beta}$ & $\boldsymbol{1/z}$ & $\boldsymbol{\alpha}= \beta z $ \\
\hline

\makecell{C-KPZ\\(initial time)} & Model Ref.~\cite{LaiSarma1991} \newline {\footnotesize \textit{($d=2+1$)}} 
& -- 
& $\dfrac{5-d}{7+d} = \dfrac{1}{5} \newline = 0.20$ 
& $\dfrac{3}{7+d} = \dfrac{3}{10} \newline = 0.30$ 
& $\dfrac{5-d}{3} = \dfrac{2}{3} \newline = 0.67$ \\
\cline{2-6}
& Experiment 
& $(\boldsymbol{\alpha_{\text{loc1}}})\newline0.71 \pm 0.03$ 
& $(\boldsymbol{\beta_{\text{1}}})\newline0.21 \pm 0.03$ 
& $(\boldsymbol{1/z_1}) \newline0.31 \pm 0.07$ 
& $(\boldsymbol{\alpha_{\text{1}}})\newline 0.68 \pm 0.18$\\
\hline
\cline{2-6}
& Simulation 
& $0.68 \pm 0.04$ 
& $0.24 \pm 0.10$ 
& $0.37 \pm 0.07$
& $0.65 \pm 0.3$\\
\hline

\makecell{TCN-KPZ\\(later time)} & Model Ref.~\cite{Song_2021} \newline {\footnotesize ($\theta \sim 0.45$)} 
& --
& 0.66 
& 0.73
& 0.92\\
\cline{2-6}
& Experiment 
& $(\boldsymbol{\alpha_{\text{loc2}}})\newline 0.87 \pm 0.07$ 
& $(\boldsymbol{\beta_{\text{2}}})\newline0.66 \pm 0.13$ 
& $(\boldsymbol{1/z_2}) \newline0.74 \pm 0.05$ 
& $(\boldsymbol{\alpha_{\text{2}}})\newline0.89 \pm 0.19$\\
\cline{2-6}
& Simulation 
& $0.77 \pm 0.04$ 
& $0.72 \pm 0.04$ 
& $0.63 \pm 0.05$  
& $1.14 \pm 0.11$\\
\hline
\end{tabular}
\caption{Comparison of theoretical, experimental, and simulated critical exponents for C-KPZ and TCN-KPZ universality classes. Theoretical values for the C-KPZ class are taken from the Lai–Das Sarma model~\cite{LaiSarma1991}, while those for the TCN-KPZ class correspond to the temporally correlated noise model~\cite{Song_2021} with $\theta \sim~0.45$. Experimental and simulation results show good agreement with their respective universality predictions, confirming the crossover from conserved to temporally correlated KPZ dynamics.}

\label{tab:1} 
\end{table}


\clearpage 

\bibliography{science_template} 
\bibliographystyle{sciencemag}


\section*{Acknowledgments}
We thank Prof.~B.~N.~Dev, Dr.~Rajeev~Rawat and Dr.~Saroj~Kumar~Nandi for valuable discussions, and Dr.~Juan~Manuel~López for insightful suggestions and comments.

\paragraph*{Funding:}
R.B. acknowledges support from the Start-up Research Grant (SRG), SERB, Govt. of India, file no. SRG/2022/000552-G. C.M. acknowledges funding from SERB (Ramanujan Fellowship, File No. RJF/2021/000129). S.R.B. is thankful to Council of Scientific and Industrial Research for grant under Emeritus Scientist Scheme (21/1170/24/EMR-II). A.K.P. acknowledges the Israel Council for Higher Education for the support of a postdoctoral fellowship through the Study in Israel program. M.H. thanks the Leona M. and Harry B. Helmsley Charitable Trust Grants No. 2018PG-ISL006 and No. 2112-04911 for support.

\paragraph*{Author contributions:}
M.M., S.G., and P.B. grew the thin films, acquired and analyzed the STM data. S.B. assisted with film growth and STM analysis. S.M. performed the MD simulations and analyzed the simulation data under the supervision of C.M. and R.B.. C.M conceived the simulations. A.K.P. synthesized the substrate crystals in M.H.’s group. A.G. and V.R.R. carried out the GIXRD measurements. R.B. conceptualized the project, performed the analysis and led the project with contributions from M.M., A.R., S.R.B., S.K., and C.M.. S.K. conceptualized the growth in simulations. R.B., M.M. S.M. and C.M. wrote the manuscript with input from all authors.

\paragraph*{Competing interests:}
There are no competing interests to declare.
\paragraph*{Data and materials availability:}
All data needed to evaluate the conclusions in the paper are present in the paper and/or the Supplementary Materials. Additional data related to this paper may be requested from the authors.


\subsection*{Supplementary materials}
Materials and Methods\\
Figs. S1 to S13\\
Movies S1 and S2


\newpage


\renewcommand{\thefigure}{S\arabic{figure}}
\renewcommand{\thetable}{S\arabic{table}}
\renewcommand{\theequation}{S\arabic{equation}}
\renewcommand{\thepage}{S\arabic{page}}
\setcounter{figure}{0}
\setcounter{table}{0}
\setcounter{equation}{0}
\setcounter{page}{1} 


\begin{center}
\section*{Supplementary Materials for\\ \scititle}


\author{Mrinal Manna$^{1\dagger}$, Sourav Mukherjee$^{1\dagger}$, Soumen Giri$^{1}$, Pramod Bhakuni$^{1}$, Sajal Barman$^{1}$, \\  Arnab Kumar Pariari$^{2}$, Anil Gome$^{1}$, Markus Hucker$^{2}$, V. Raghavendra Reddy$^{1}$, Anupam Roy $^{3}$,\\ Sudipta Roy Barman$^{1}$, Smarajit Karmakar$^{4}$, Chandana Mondal$^{1}$, Rajib Batabyal$^{1\star}$\\ 
\small$^\ast$Corresponding author. Email: rajibbata@csr.res.in\\
\small$^\dagger$These authors contributed equally to this work.}
\end{center}

\subsubsection*{This PDF file includes:}
Materials and Methods\\
Figures S1 to S13\\
Captions for Movies S1 to S2 

\subsubsection*{Other Supplementary Materials for this manuscript:}
Movie S1 \\
Movie S2 

\newpage


\subsection*{Materials and Methods}

\subsection*{Material details}

\textbf{(a) Substrate:}  
The $30\%$ Sb-doped MnBi$_2$Te$_4$ (MBST) single crystals were grown using the flux method. To synthesize an $n\%$ Sb-doped MBST crystal, high-purity ($99.999\%$) Mn powder, Bi shots, Te chunks, and Sb shots were mixed in a molar ratio of Mn:Sb:Bi:Te = 1 : 10$\times$ n : 10$\times$(1 - n) : 16 placed in an alumina crucible of a Canfield crucible set. The assembly was sealed in an evacuated silica ampule under a pressure below $10^{-5}$~mbar. The ampule was heated to 1173~K over two days, held for 12~h, and then slowly cooled to 878~K at a rate of 10~K~h$^{-1}$. In the final stage, the temperature was lowered from 878~K to 863~K over 336~h and maintained for 72~h before centrifugation at 3000~rpm to remove the residual flux. Shiny, plate-like single crystals were obtained from the bottom crucible. The Sb concentration in our samples was determined to be $\sim30\%$~\cite{sajal2024 }.\\
\textbf{(b) Interface Growth:}  
Sn (99.99\%) was deposited on MBST surfaces using a water-cooled Knudsen cell~\cite{shukla2004} at a cell temperature of 1143~K, while the sample was maintained at room temperature ($\sim$303~K). The deposition rate was fixed at 0.83~ML~min$^{-1}$ during all deposition cycles. The chamber pressure remained at $\sim 2 \times 10^{-10}$~mbar throughout the growth.

\subsection*{Measurement details}

Atomically clean MBST(0001) surfaces were prepared \textit{in situ} by the scotch-tape exfoliation method inside an ultrahigh-vacuum (UHV) chamber with a base pressure of $1.0 \times 10^{-10}$~mbar. Scanning tunneling microscopy (STM) measurements were performed using a Scienta Omicron GmbH variable-temperature STM system maintained at $\sim 4 \times 10^{-11}$~mbar throughout the experiments and operated at room temperature.  
Auger electron spectroscopy (AES) was carried out immediately after each adlayer deposition cycle. A chemically etched tungsten (W) tip was used for all STM measurements. Grazing-incidence X-ray diffraction (GIXRD) measurements were performed at ambient temperature and pressure.

\subsection*{Simulation details}

Molecular dynamics (MD) simulations were performed to model Sn deposition on two substrates corresponding to distinct experimental regimes.

\textbf{Regime I:} Sn was deposited on a Bi$_2$Te$_3$ slab comprising a single quintuple layer (QL) with a Te–Bi–Te–Bi–Te stacking sequence, having a thickness of approximately 1~nm and lateral dimensions of $\approx$ 32~nm. Bi$_2$Te$_3$ was chosen as an analog of the experimentally used MBST-based substrate, since the latter contains magnetic Mn and Sb dopants for which reliable empirical potentials are unavailable. To mimic bulk anchoring, the bottom atomic layers of the substrate were fixed, while the upper layers were allowed to relax dynamically.

\textbf{Regime II:} In this case, Sn atoms were deposited on a buckled honeycomb Sn layer, representing a stanene sheet with a buckling height of 0.085~nm and a lattice constant of 0.467~nm. This model represents an idealized free-standing stanene layer, serving as a reference for strain-free growth conditions.

Sn was deposited on a single buckled honeycomb Sn layer representing a stanene monolayer.

In both regimes, periodic boundary conditions were applied along the lateral ($x$, $y$) directions, and free boundaries were used along the growth direction ($z$). The simulation box extended up to 20~nm in the $z$-direction—sufficient to eliminate interactions with periodic images.

Interatomic interactions among Bi, Te, and Sn atoms were modeled using the Elliott–Akerson (2015) universal Lennard–Jones (LJ) potential distributed via OpenKIM~\cite{OpenKIM-MO:959249795837:003}. The pair potential between atoms $i$ and $j$ separated by a distance $r$ is given by
\begin{equation}
    V_{ij}(r) = 4\epsilon_{ij}\left[ \left( \frac{\sigma_{ij}}{r} \right)^{12} - \left( \frac{\sigma_{ij}}{r} \right)^{6} \right],
\end{equation}
where $\epsilon_{ij}$ denotes the potential well depth and $\sigma_{ij}$ the distance at which the potential vanishes. Cross-interactions were computed using Lorentz–Berthelot mixing rules. While this LJ form captures adsorption and diffusion behavior, it does not include directional bonding, charge transfer, or spin–orbit effects; thus, our results focus on morphological evolution and universality-class identification rather than electronic or magnetic properties.

Sn atoms were introduced sequentially above the instantaneous surface at a flux of one atom per ten MD steps, corresponding to an experimental deposition rate of $\approx$0.83~ML~min$^{-1}$ when mapped to MD timescales. One monolayer (ML) was defined as the number of Sn atoms required to completely cover the topmost substrate surface. All simulations were performed in LAMMPS ~\cite{LAMMPS} using the \textit{metal} unit system with a timestep of 1~fs. The neighbor list was updated every timestep with a skin distance of 0.2~nm. During deposition, the incoming atoms were maintained at 1143~K to emulate the experimental Knudsen-cell temperature, while the substrate was held at 300~K. This setup ensured that the incoming atoms acquired realistic kinetic energies consistent with the experimental flux, while the substrate remained thermally stable. After each deposition cycle, the system was cooled to 300~K over 1~ns and equilibrated for an additional 1~ns. Atomic configurations, potential energies, and per-atom stresses were recorded every 5,000 steps. Density profiles along the growth direction were extracted to monitor film thickness and surface roughness. Atomic trajectories were visualized and analyzed using OVITO ~\cite{ovito} to generate structural snapshots and perform qualitative post-processing of the evolving morphology.

\newpage











\begin{figure} 
	\centering
	\includegraphics[width=1\textwidth]{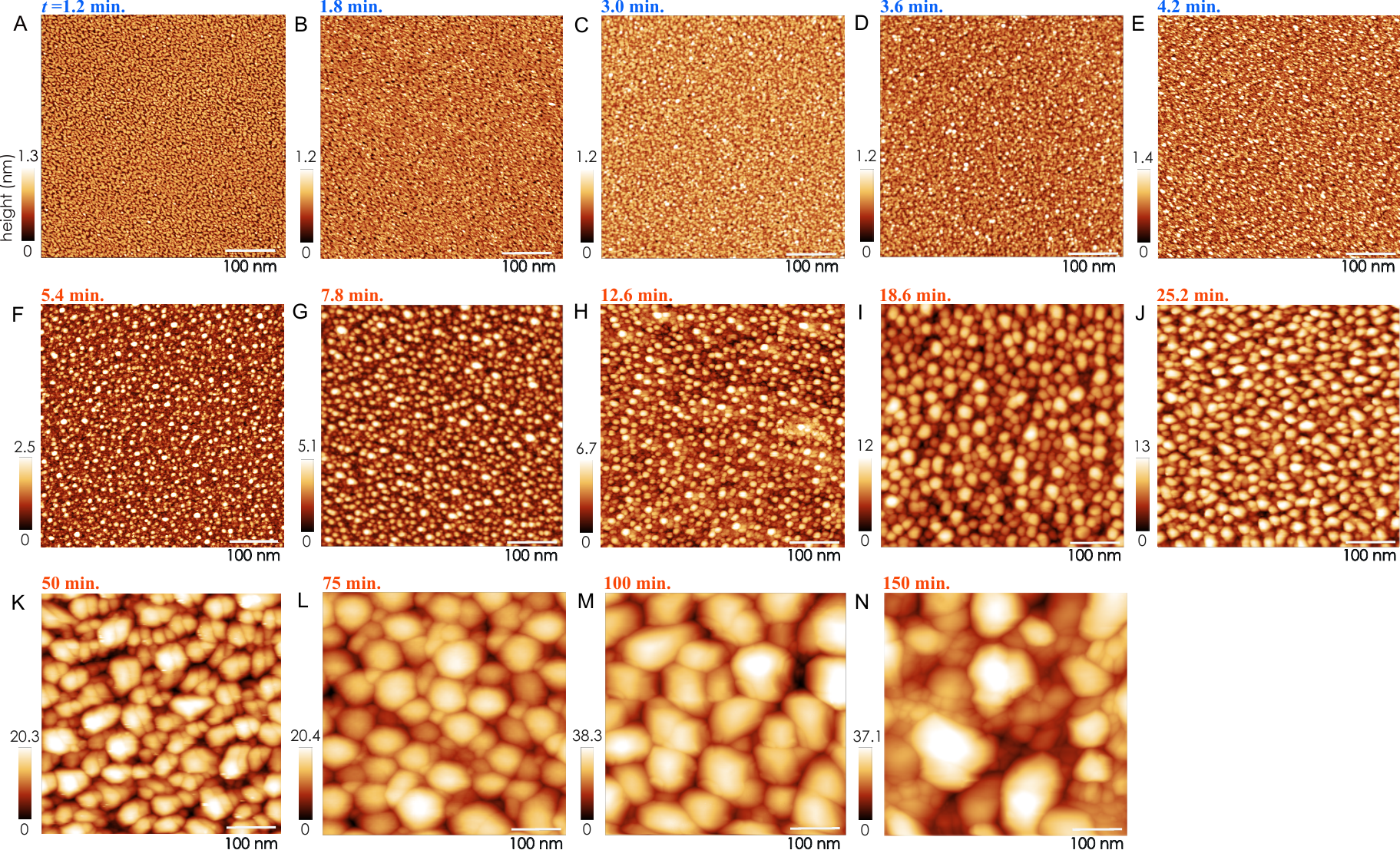} 

	\caption{\textbf{(A–N)} Wide-area ($500~\mathrm{nm} \times 500~\mathrm{nm}$) STM topographies at successive growth times showing the morphological evolution of Sn films from initial island nucleation to clustered mounds and faceted grains.}
	\label{figs1} 
\end{figure}

\begin{figure} 
	\centering
	\includegraphics[width=1.0\textwidth]{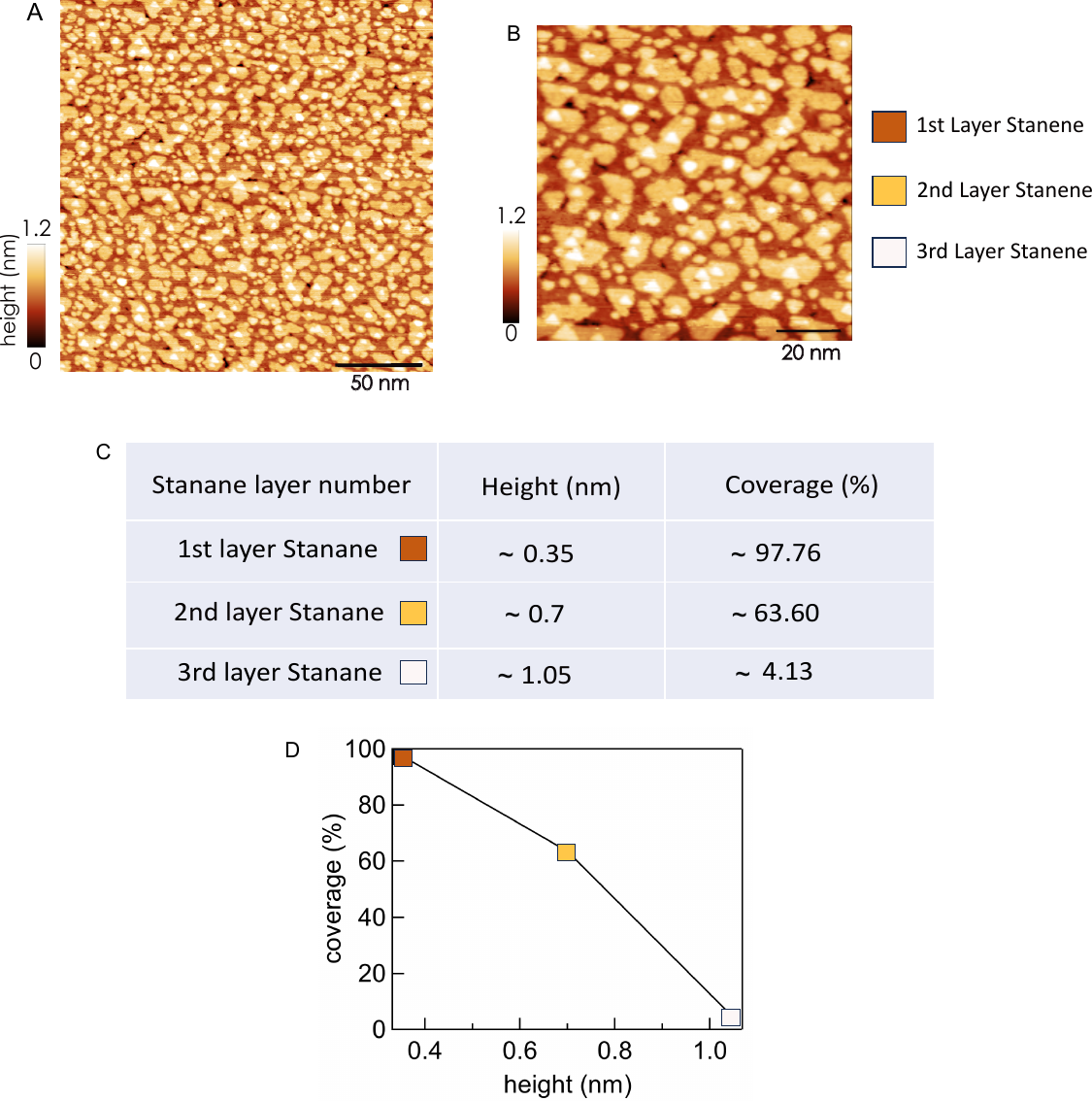} 
	\caption{\textbf{(A)} STM topography ($220~\mathrm{nm} \times 220~\mathrm{nm}$) of the stanene island region. \textbf{(B)} Zoomed view ($98~\mathrm{nm} \times 98~\mathrm{nm}$) showing successive stanene layers (first to third). \textbf{(C)} Table listing layer heights and lateral coverage fractions. \textbf{(D)} Comparison of lateral area coverage for different stanene layers, revealing progressive layer filling during growth.}

	\label{figs2} 
\end{figure}

\begin{figure} 
	\centering
	\includegraphics[width=0.7\textwidth]{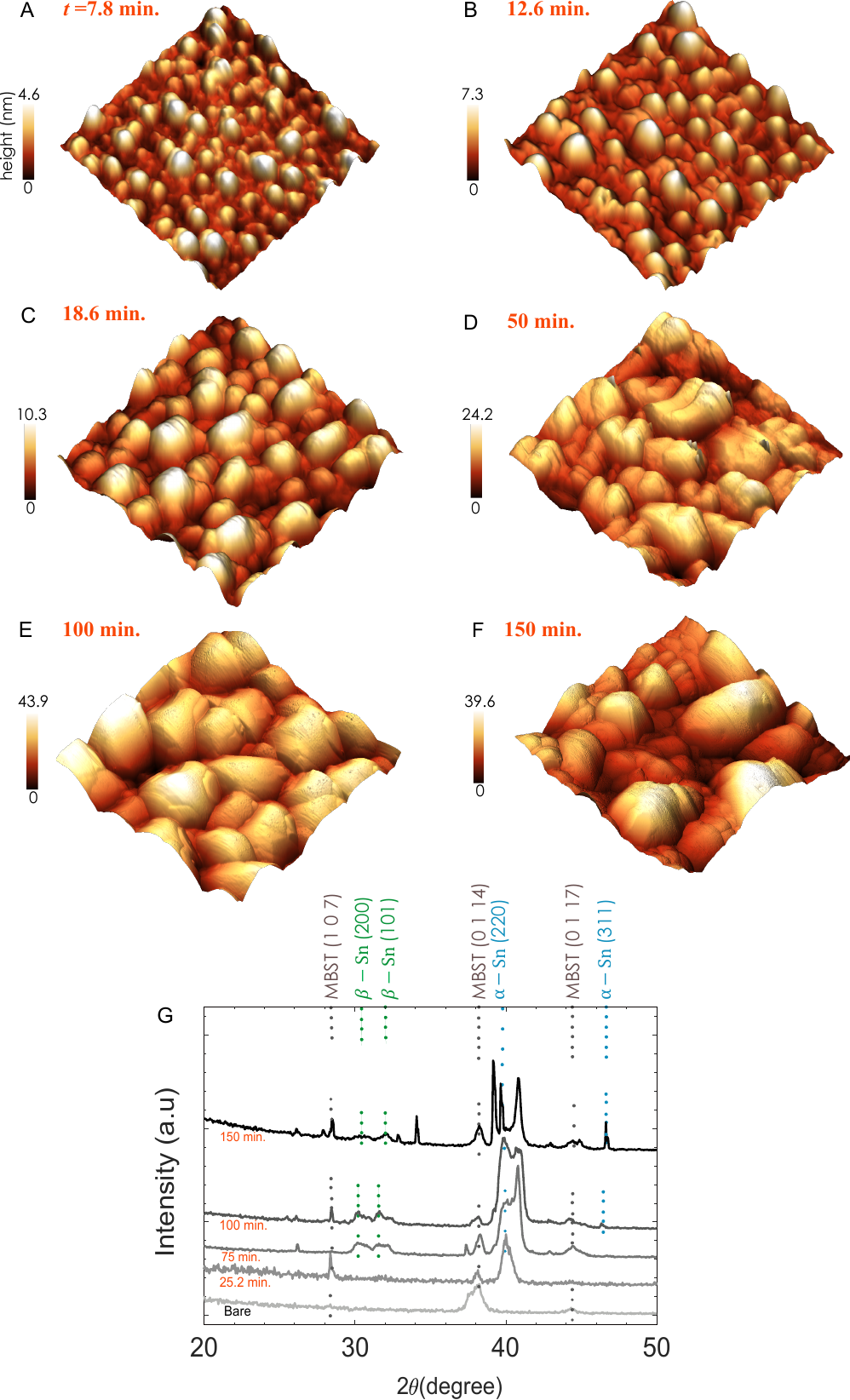} 

	\caption{\textbf{(A–F)} 3D STM topographies within the temporally correlated noise regime illustrating the evolution from mound-like structures to faceted grains. \textbf{(G)} Grazing-incidence X-ray diffraction (GIXRD) data acquired at successive deposition times confirm the emergence of mixed $\alpha$- and $\beta$-Sn phases, following an initial $\alpha$-Sn–dominated growth.}

	\label{figs3} 
\end{figure}

\begin{figure} 
	\centering
	\includegraphics[width=1\textwidth]{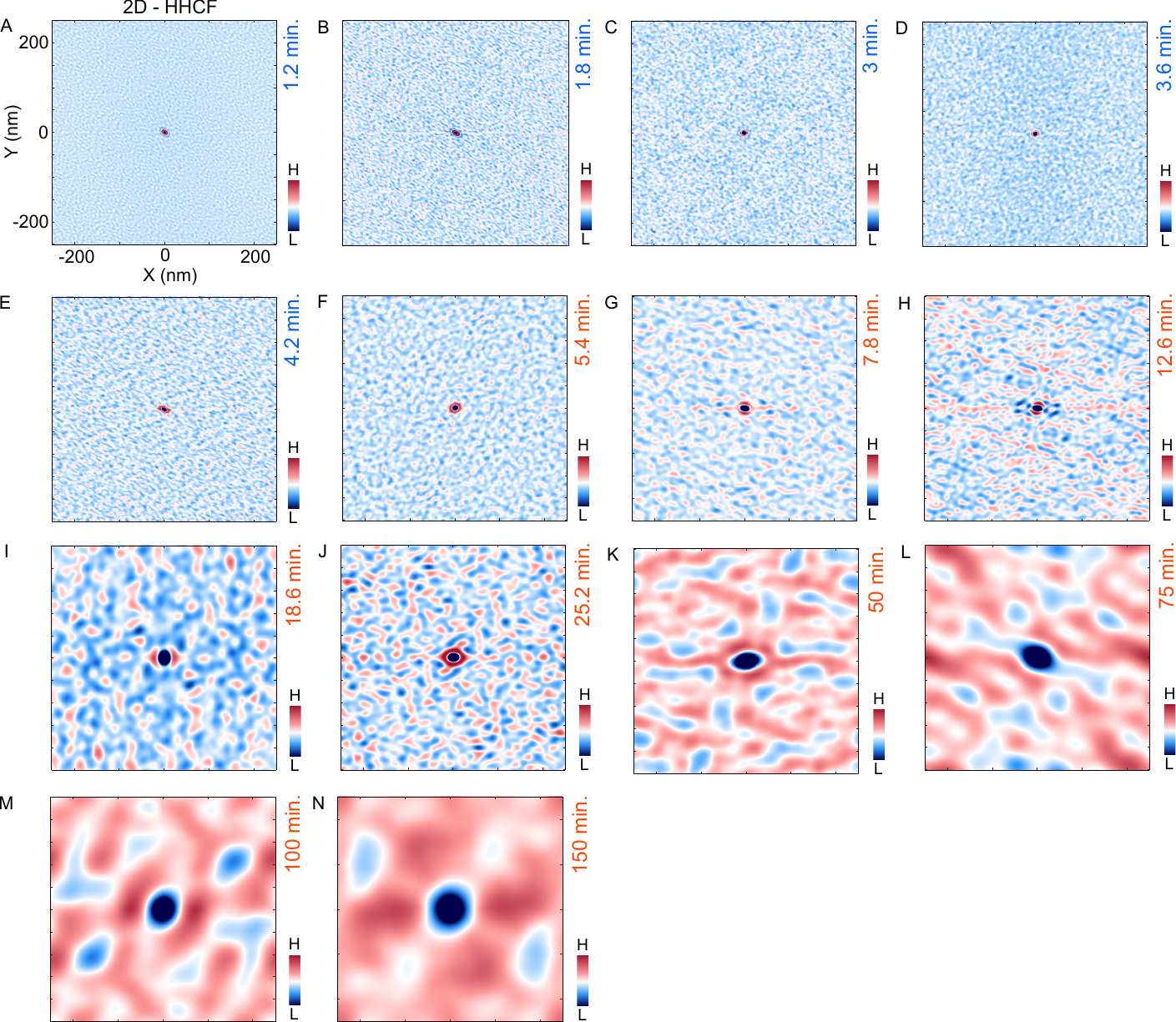} 

	\caption{\textbf{(A–N)} Temporal evolution of the 2D height–height correlation function (HHCF) during Sn growth. All panels share identical $x$–$y$ scales; color indicates relative correlation amplitude. Early-stage maps (\textbf{A–E}) show isotropic, low-amplitude correlations corresponding to smooth films, while later maps (\textbf{F–N}) display anisotropic, extended features reflecting the onset of rough surface morphology.}

	\label{figs4} 
\end{figure}

\begin{figure} 
	\centering
	\includegraphics[width=1\textwidth]{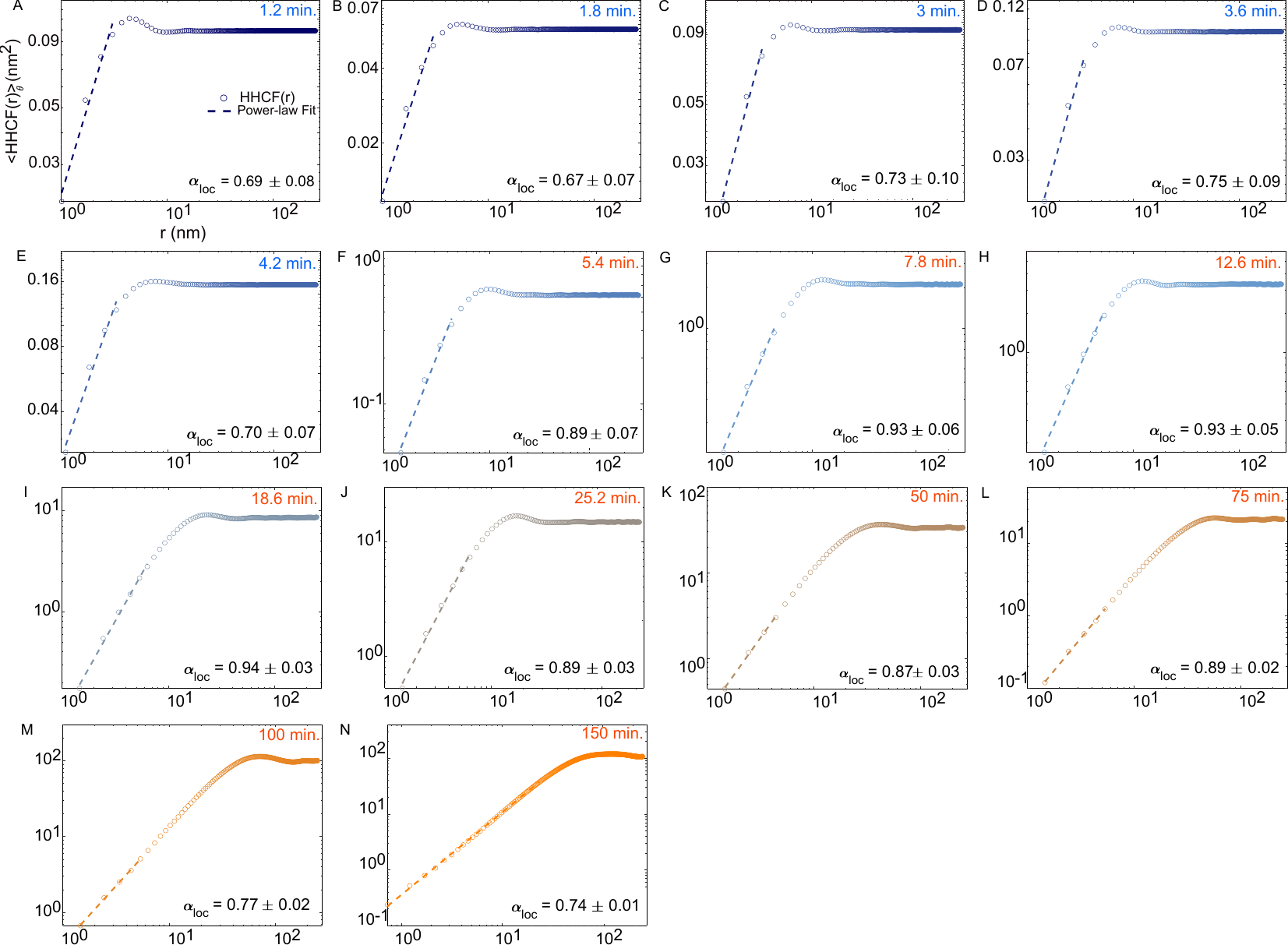} 

	\caption{\textbf{(A–N)} Log–log plots of the angularly averaged height–height correlation function $\langle H(r,t) \rangle_{\theta}$ as a function of spatial separation $r$ for different growth times. Experimental data are fitted with a power-law function, \( \langle H(r,t) \rangle_{\theta} \sim \rho^2 r^{2\alpha_{\mathrm{loc}}} \) (dashed lines), for \( r \ll \xi(t) \) to extract the local roughness exponent $\alpha_{\mathrm{loc}}$, indicated in each panel.}

	\label{figs5} 
\end{figure}

\begin{figure} 
	\centering
	\includegraphics[width=1\textwidth]{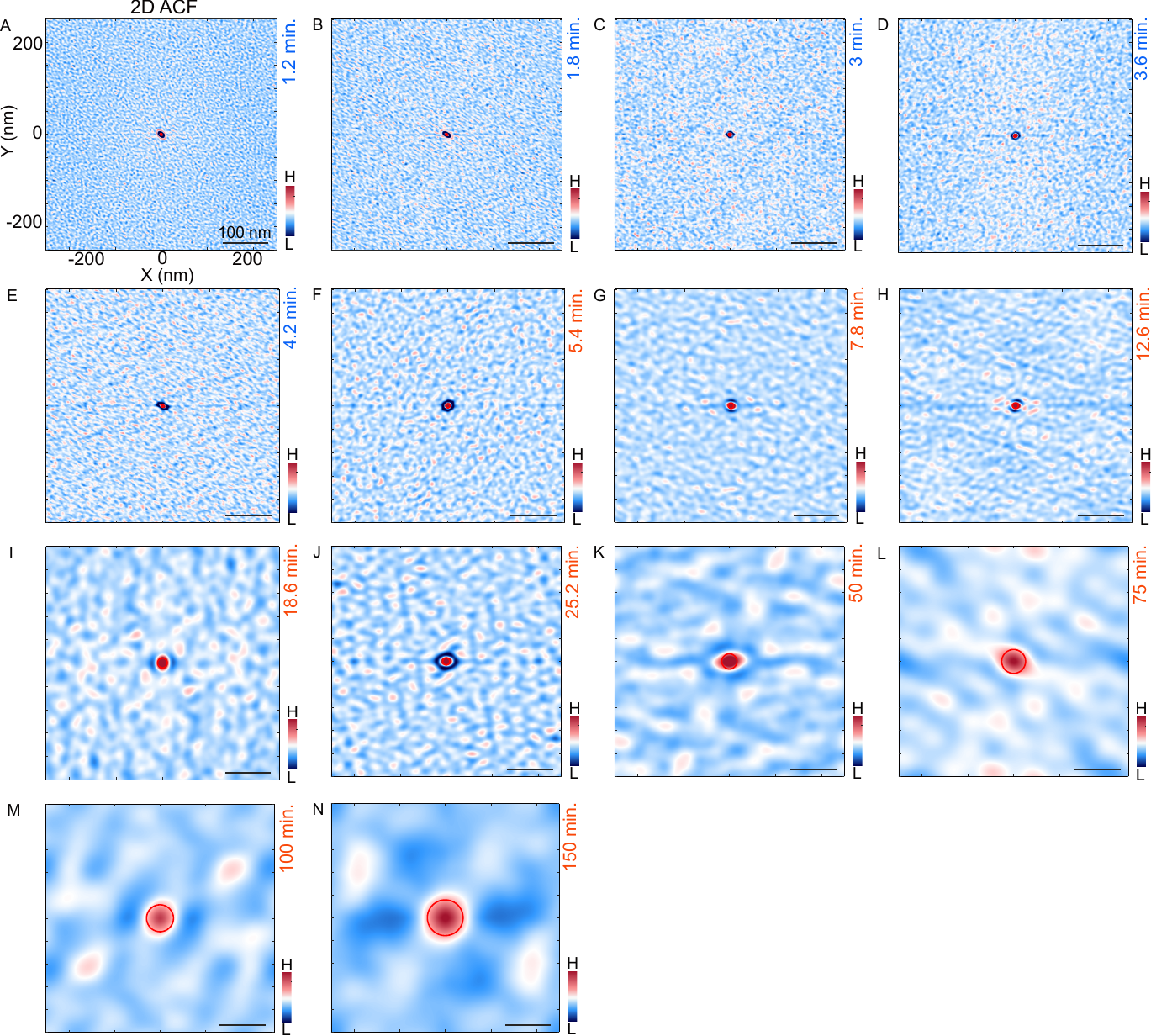} 

	\caption{\textbf{(A–N)} Two-dimensional autocorrelation functions (ACFs) derived from STM topographies at different growth times. All plots share the same spatial scale bar (100~nm), and the color bar indicates relative correlation amplitudes. The red circle marks the radius where the ACF decays to $1/e$ of its maximum value. Early-stage maps (\textbf{A–E}) retain fine structural details, reflecting short-range surface order, while at later times (\textbf{F–N}), extended correlations emerge, indicating the development of long-range surface order.}

	\label{figs6} 
\end{figure}

\begin{figure} 
	\centering
	\includegraphics[width=1\textwidth]{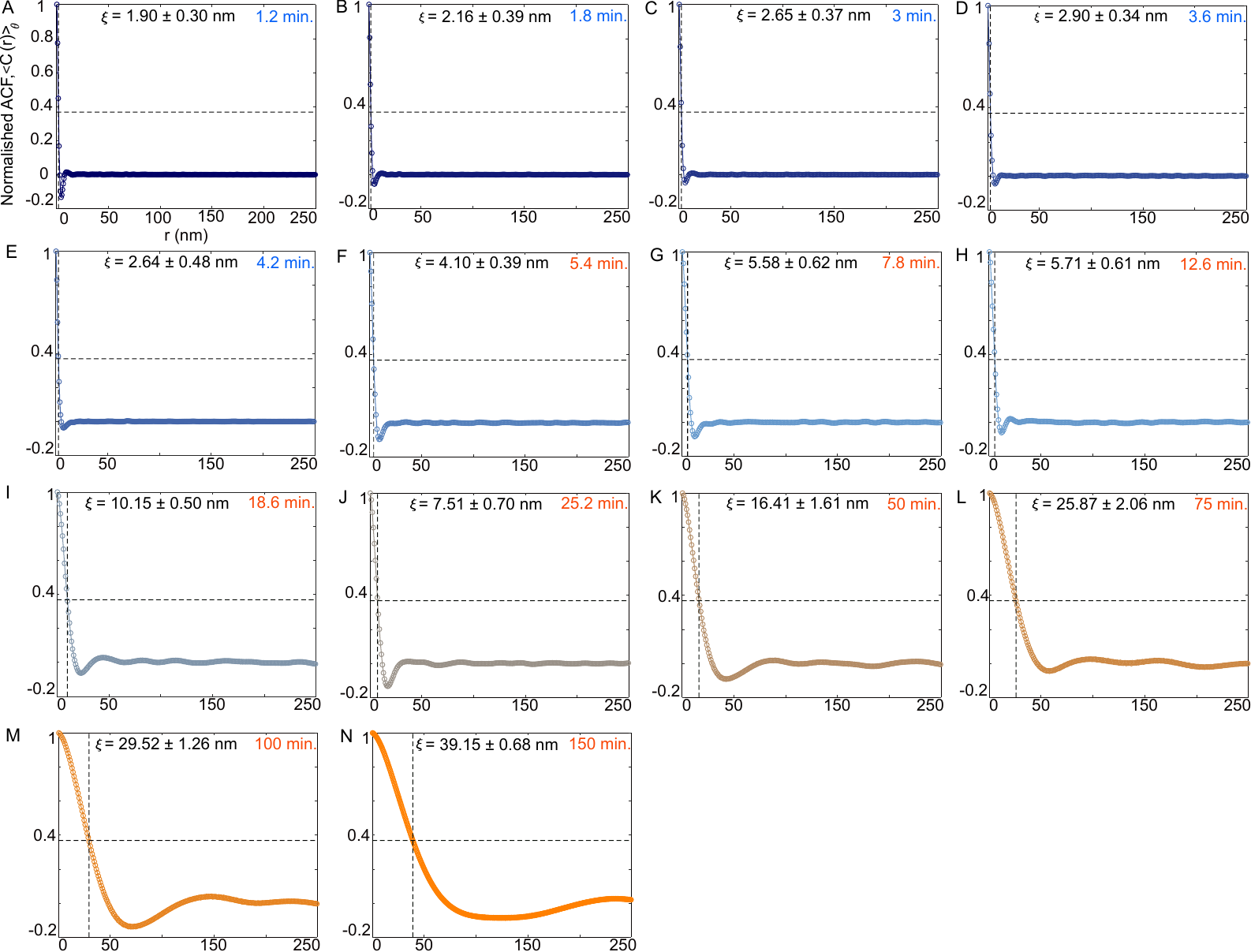} 

	\caption{\textbf{(A–N)} Normalized angle-averaged autocorrelation functions, $\langle C(r,t) \rangle_{\theta}$, plotted as a function of spatial distance $r$ for different growth times. The correlation length $\xi$ is extracted from the intersection point where the normalized $\langle C(r,t) \rangle$ decays to $1/e$ of its maximum value (indicated by the crossing of the dotted horizontal and vertical lines).}

	\label{figs7} 
\end{figure}

\begin{figure} 
	\centering
	\includegraphics[width=0.7\textwidth]{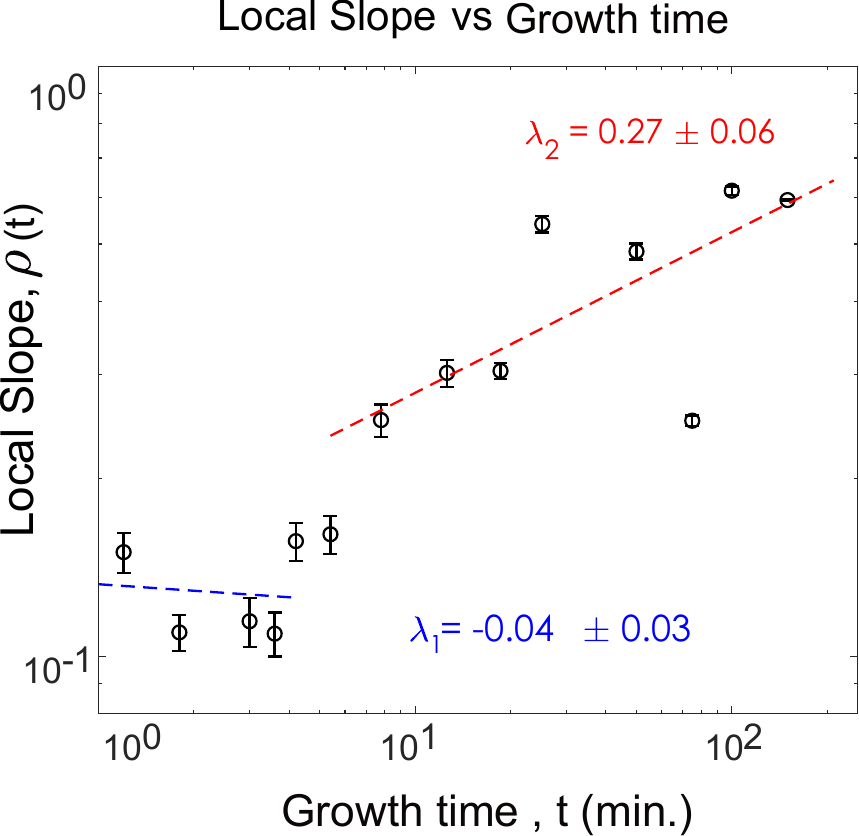} 

	\caption{Plot of the local surface slope, $\rho(t)$, as a function of growth time. The data are fitted with a power-law relation, $\rho(t) \sim t^{\lambda}$, to extract the slope exponent $\lambda$.}

	\label{figs8} 
\end{figure}

\begin{figure} 
	\centering
	\includegraphics[width=0.7\textwidth]{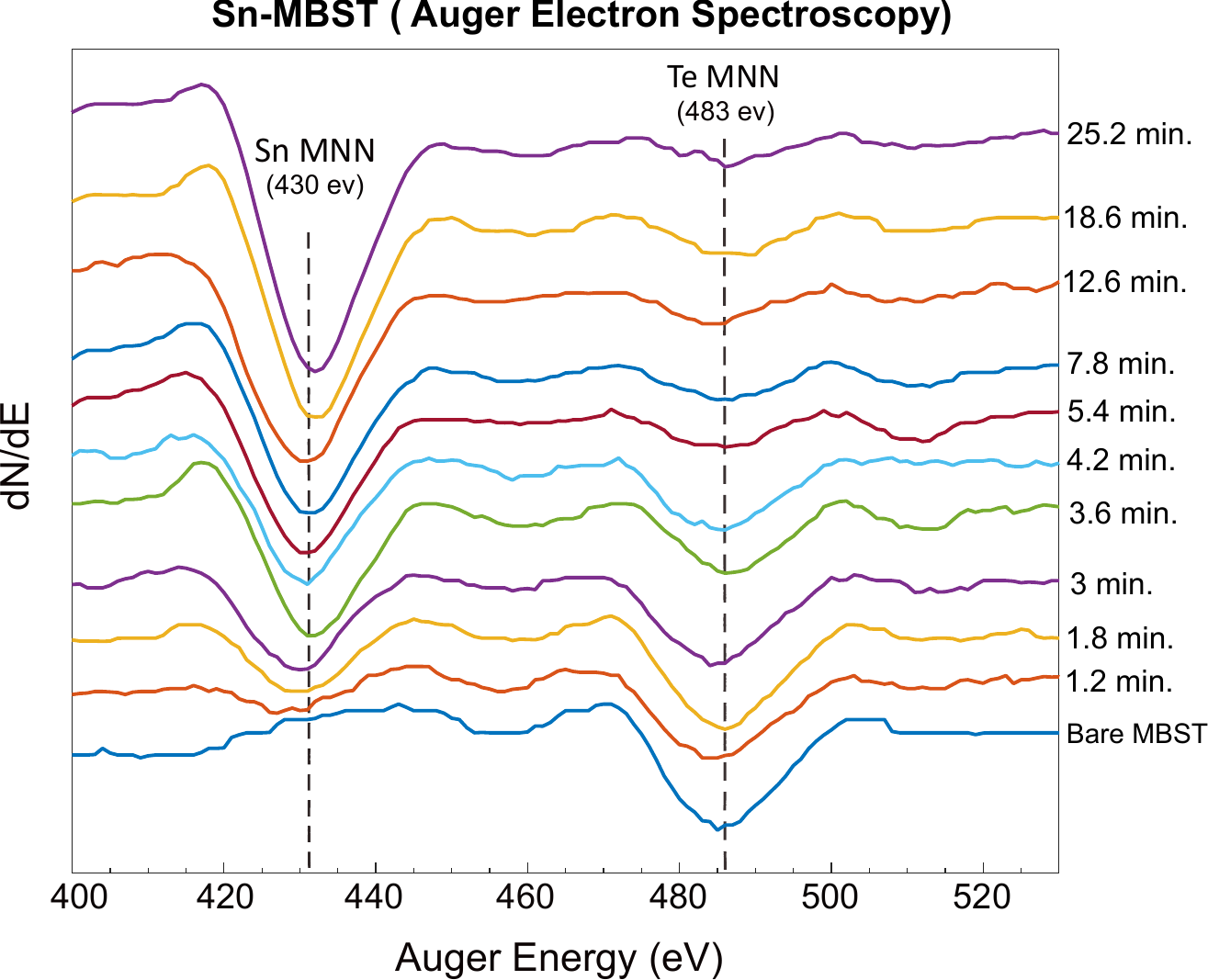} 

	\caption{Auger electron spectroscopy (AES) measured as a function of deposition time, highlighting the Sn MNN (430~eV) and Te MNN (483~eV) peaks. Spectra are staggered along the vertical axis for clarity.}
    
	\label{figs9} 
\end{figure}

\begin{figure} 
	\centering
	\includegraphics[width=1\textwidth]{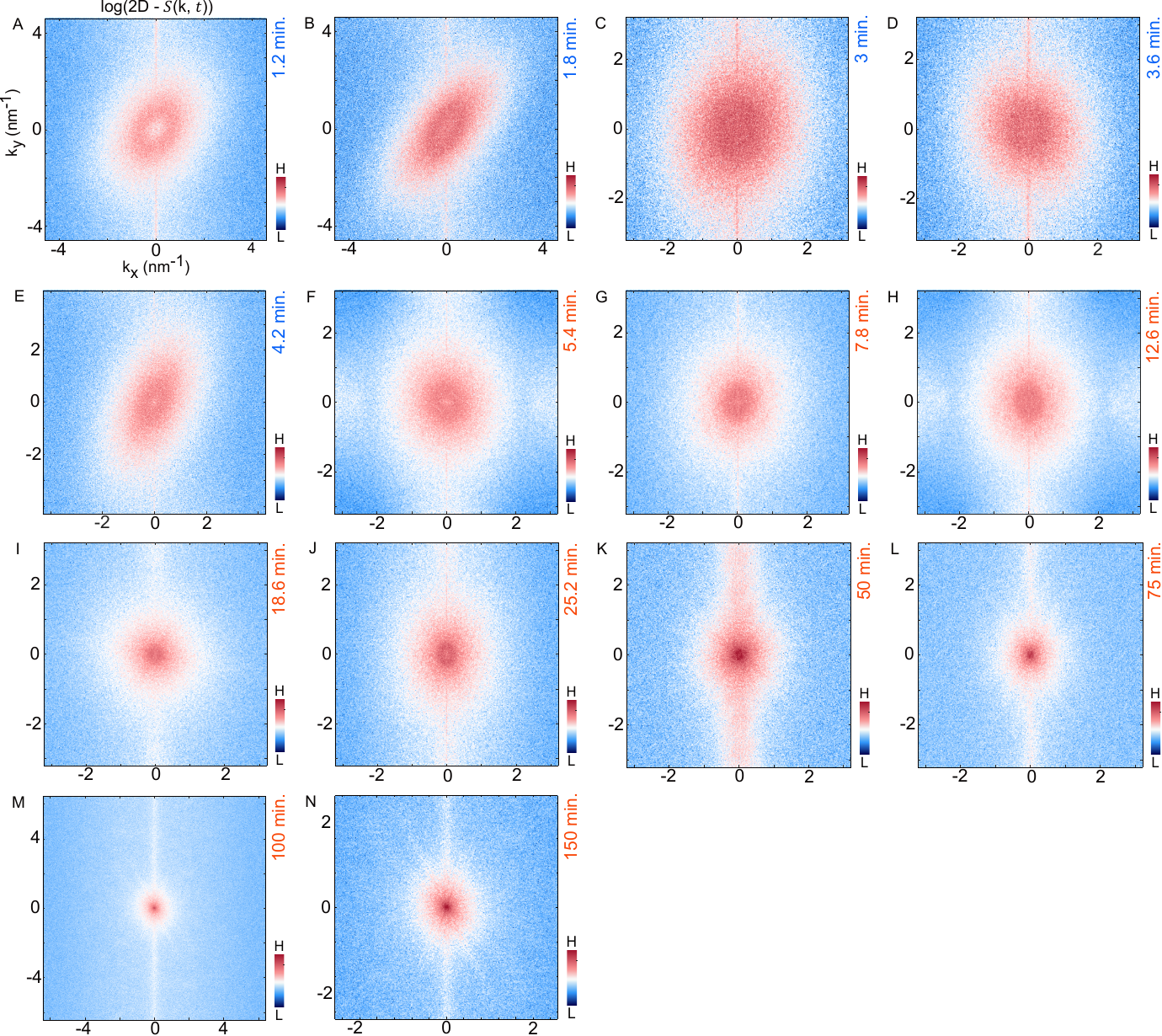} 

	\caption{\textbf{(A–N)} Logarithmic plots of the two-dimensional structure factor, $S(\mathbf{k},t)$, at different deposition times. With increasing growth time, the intensity distribution narrows, reflecting a loss of fine structural details and the emergence of broader surface motifs. Color bars indicate normalized spectral intensity (H: high, L: low), and axes are in reciprocal space ($k_x$, $k_y$) in units of nm$^{-1}$.}

	\label{figs10} 
\end{figure}

\begin{figure} 
	\centering
	\includegraphics[width=1\textwidth]{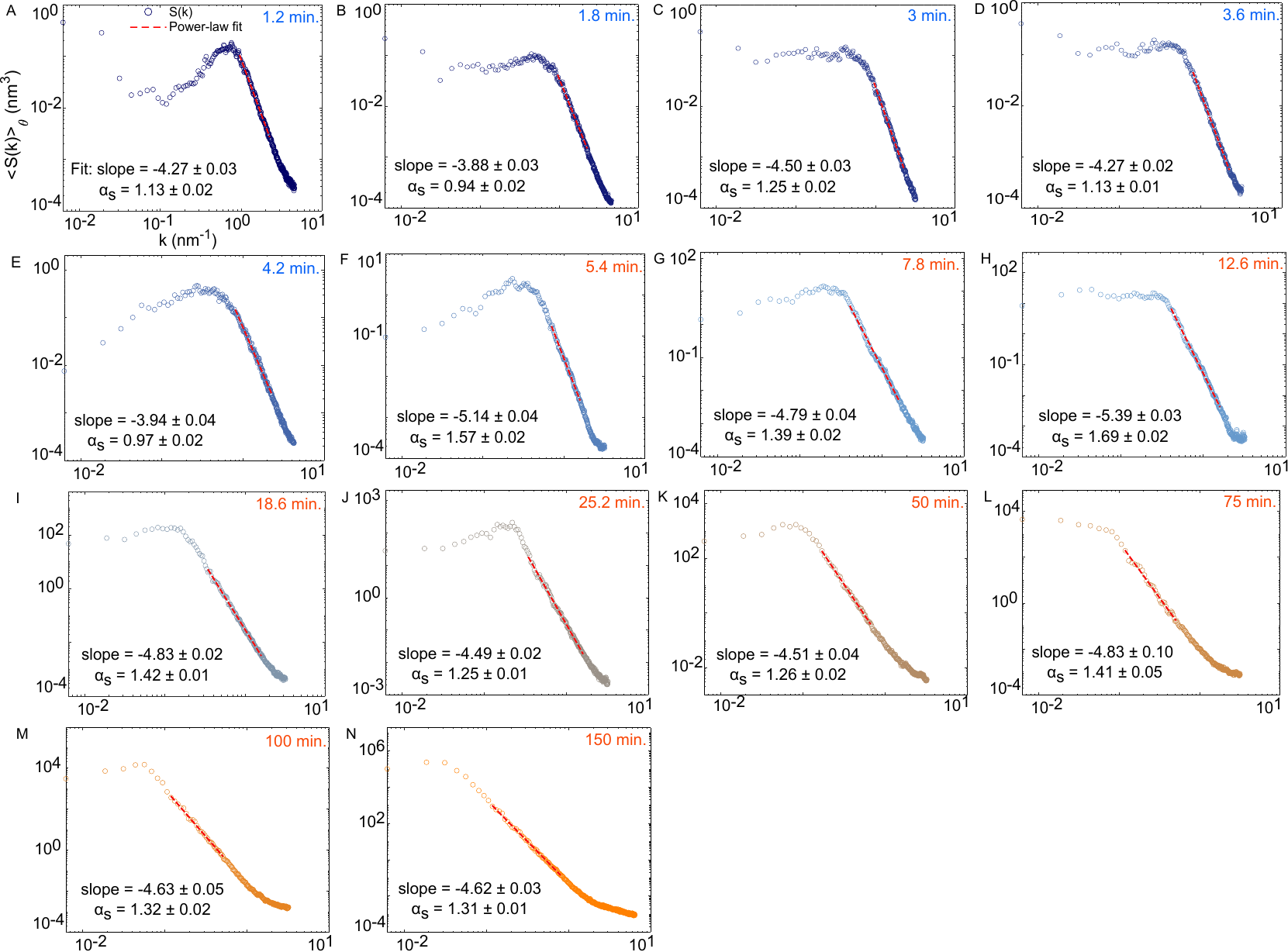} 

	\caption{\textbf{(A–N)} Log–log plots of the angularly averaged structure factor, $\langle S(k,t) \rangle_{\theta}$, as a function of spatial frequency $k$ for STM topographies. The  linear scaling regime is fitted with a power-law, $S(k) \sim k^{-(2\alpha_s+2)}$ (red dotted lines). Extracted slopes and the corresponding spectral roughness exponents, $\alpha_s$, are indicated within each panel.}

	\label{figs11} 
\end{figure}

\begin{figure} 
	\centering
	\includegraphics[width=1\textwidth]{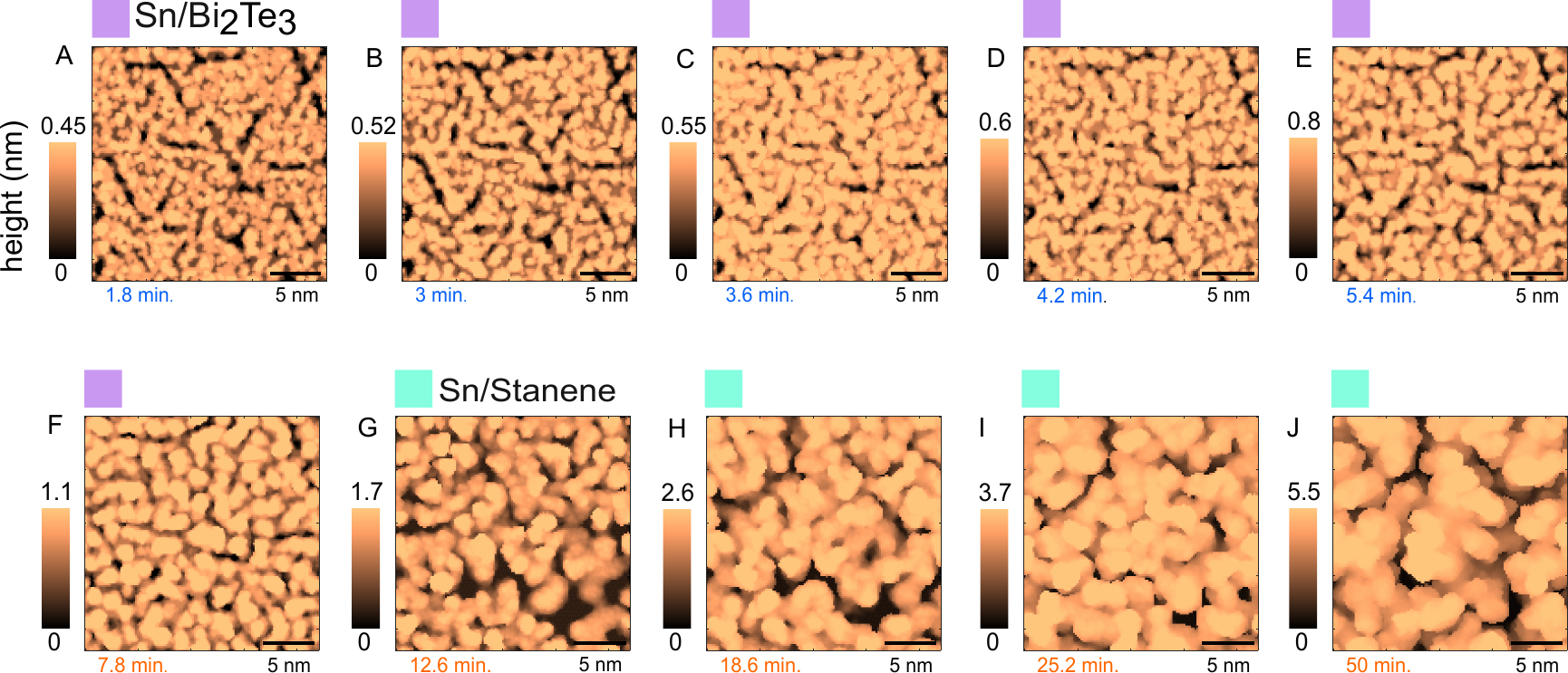} 

	\caption{(\textbf{A–J}) Simulated surface topographies of Sn growth at different deposition times. (\textbf{A–F}) Growth on a Bi$_2$Te$_3$ substrate. (\textbf{G–J}) Continued growth on a stanene substrate .}

	\label{figs12} 
\end{figure}

\begin{figure} 
	\centering
	\includegraphics[width=1\textwidth]{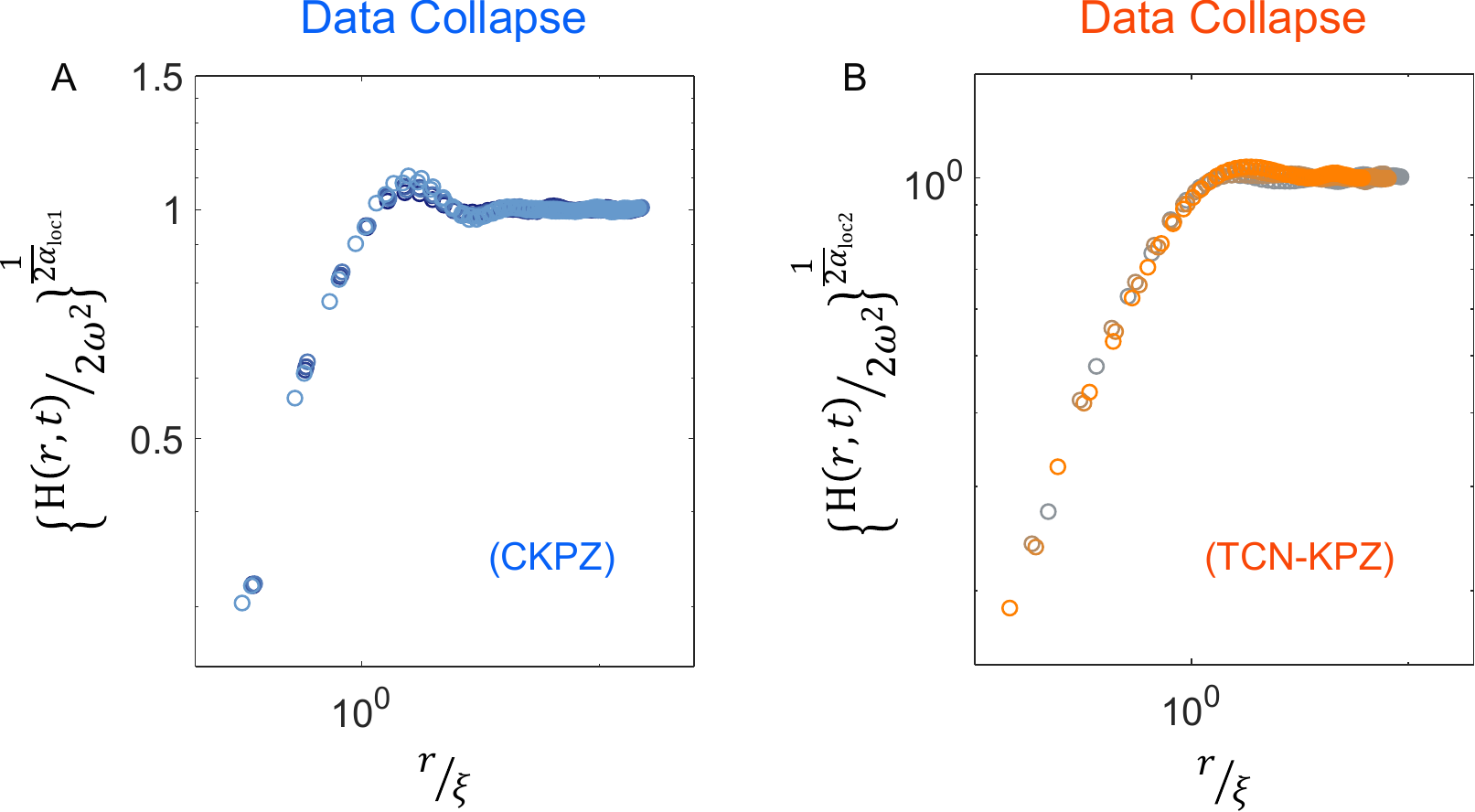} 

	\caption{(\textbf{A–B}) Data collapse of the angularly averaged height–height correlation function, $\langle H(r,t) \rangle_{\theta}$, obtained from simulated Sn surface topographies in both growth regimes.}

	\label{figs13} 
\end{figure}

%


\clearpage 

\paragraph{Caption for Movie S1.}
\textbf{Sn growth on a Bi$_2$Te$_3$} Molecular dynamics simulation of Sn growth on a Bi$_2$Te$_3$ substrate showing the evolution surface morphologies.

\paragraph{Caption for Movie S2.}
\textbf{Sn growth on a Stanene} Molecular dynamics simulation of Sn growth on a stanene substrate with growth time.



\end{document}